\documentclass[showpacs,aps,floats,floatfix,12pt,letterpaper]{revtex4}

\usepackage{multirow} 
\usepackage{rotating} 
\usepackage{graphicx} 
\usepackage{tabularx} 
\usepackage{axodraw}  %
\usepackage{pstricks}
\usepackage{amssymb}

\newcommand{\beq}{\begin{equation}} 
\newcommand{\eeq}{\end{equation}} 
\newcommand{\beqa}{\begin{eqnarray}} 
\newcommand{\eeqa}{\end{eqnarray}} 
\newcommand{\mx}{\left[\begin{array}} 
\newcommand{\finmx}{\end{array}\right]} 
\newcommand{\mxp}{\left(\begin{array}} 
\newcommand{\finmxp}{\end{array}\right)} 
\newcommand{\casos}{\left\{\begin{array}} 
\newcommand{\fincasos}{\end{array}\right.} 
\newcommand{\rcasos}{\left.\begin{array}} 
\newcommand{\rfincasos}{\end{array}\right\}} 
\newcommand{\mean}[1]{{\left<#1\right>}}

\newcommand{\lsim}{\lesssim}
\newcommand{\gsim}{\gtrsim}
\def\ominus#1{\ \rlap{\raise 0pt \hbox{$#1$}}{\raise 1pt \hbox{$^{^{(-)}}$}}\ }

\renewcommand{\a}{\alpha}

\newcommand{\s}{\sigma}

\newcommand{\cerenkov}{\v{Cerenkov}\ }

\newcommand{\lik}{{\cal L}}

\renewcommand{\ae}{{\overline{\nu}_e}}
\renewcommand{\ae}{{\overline{\nu}_e}}

\newcommand{\barr}[1]{\overline{#1}}

\newcommand{\blankline}[1]{\multicolumn{#1}{c}{\strut}\\}

\begin{document}


\title{   
  %
Constraints on neutrino masses from 
a Galactic supernova neutrino signal 
at present and future detectors
}


\author{Enrico Nardi\footnote{
enrico.nardi@lnf.infn.it 
}$^{a,b}$ and
Jorge I. Zuluaga\footnote{jzuluaga@naima.udea.edu.co}$^b$}


\affiliation{
$^a$INFN, Laboratori Nazionali di Frascati, C.P. 13, I00044 Frascati,  Italy \\
$^b$Instituto de F\'\i sica,  Universidad de Antioquia, A.A.{\it 1226}, Medell\'\i n, Colombia}



%
\begin{abstract}
\noindent
We study the constraints on neutrino masses that could be derived from
the observation of a Galactic supernova neutrino signal with present
and future neutrino detectors.  Our analysis is based on a recently
proposed method that uses the full statistics of neutrino events and
does not depend on particular astrophysical assumptions. The
statistical approach, originally justified mainly in terms of
intuitive reasoning, is put on a more solid basis by means of Bayesian
inference reasoning. Theoretical uncertainties in the neutrino signal
time profiles are estimated by applying the method to two widely
different supernova models.  Present detectors can reach a sensitivity
down to 1 eV. This is better than limits from tritium $\beta$-decay
experiments, competitive with the most conservative results from
neutrinoless double $\beta$-decay, less precise but less dependent
from prior assumptions than cosmological bounds.  Future megaton water
\v{C}erencov detectors will allow for about a factor of two
improvement. However, they will not be competitive with the next
generation of laboratory experiments.
\bigskip 


%
\end{abstract} 

\bigskip

\pacs{14.60.Pq, 97.60.Bw\\
{\it Key words:} Supernova neutrinos; neutrino mass.\\
Published in {\bf Nucl.Phys.B731:140-163,2005} 
} 


\maketitle




\section{Introduction}
\label{sec:introduction}
During the past few years, atmospheric \cite{Atmospheric}
and solar \cite{Solar,Ahmad:2001an} 
neutrino experiments provided strong evidences for neutrino flavor
oscillations and therefore for non vanishing neutrino masses.  The
KamLAND results \cite{KamLANDResults} 
on the depletion of the $\bar \nu_e$ flux from nuclear power plants in
Japan, and the K2K indication of a reduction in the 
$\nu_\mu$ flux from the KEK accelerator, gave a final confirmation of this picture.
However, to date all the evidences for neutrino masses come from
oscillation experiments, that are only sensitive to mass square
differences and cannot give any informations on single mass values.
The challenge of measuring the absolute value of neutrino masses is
presently being addressed by means of a remarkably large number of
different approaches, ranging from laboratory experiments to a
plethora of methods that relay on astrophysical and cosmological
considerations (for recent reviews see
\cite{Paes:2001nd,Bilenky:2002aw}).  From the study of the end-point
of the electron spectrum in tritium $\beta$-decay, laboratory
experiments have set the limit $m_{\nu_e}< 2.2\,$eV
\cite{Tritium}. If neutrinos are Majorana particles, the non observation of
neutrinoless double $\beta$ decay can constrain a particular combination of
the three neutrino masses.  Interpretation of these experimental results is
affected by theoretical uncertainties related to nuclear matrix elements
calculations, and this reflects in some model dependence of the corresponding
limits, that lie in the range $m_\nu^{\rm eff} < 0.2 - 1.3\,$eV
\cite{Neutrinoless,Bilenky:2002aw}.  Tight bounds $\sum_i m_{\nu_i} <
0.6 - 1.8\,$eV have been recently set using WMAP observations of
cosmic microwave background anisotropies, galaxies redshift surveys
and other cosmological data (for a recent review see
\cite{Hannestad:2004nb} and references therein).  However, these
limits become much looser if the set of assumptions on which they rely
is relaxed (see \cite{Cosmology} for discussions on this point). 
For example, by relaxing the hypothesis that the spectrum of CMB
fluctuations is described by a single power law, consistent
cosmological models have been constructed in which the neutrino masses
can be of order eV \cite{Blanchard:2003du}.  Cosmological constraints
on neutrino masses might even be completely evaded in exotic scenarios
where neutrinos annihilate into hypothetical light bosons,
implying a suppression of their contribution to the cosmic matter
density and negligible effects on structure formation at large scales
\cite{Beacom:2004yd}.

As it was realized long time ago, valuable informations on the neutrino
masses could also be provided  by the detection of neutrinos from a Supernova
(SN) explosion \cite{SupernovaSeminal}. 
The basic idea relies on the time-of-flight delay $\Delta t$
that a neutrino of mass $m_\nu$ and energy $E_\nu$ traveling a
distance $L$ would suffer with respect to a massless particle:
\beq
\Delta t = \frac{L}{v} - L 
\approx 
5.1\,{\rm ms}\;\left(
\frac{L}{10\,\rm kpc}\right)  
\! \left(\frac{10\, \rm
    MeV}{E_\nu}\right)^2  \!\! 
  \! \left(\frac{m_\nu}{1\,\rm eV}\right)^2\,.
\label{eq:delay}
\eeq
Indeed, already in the past the detection of about two dozens of
neutrinos from SN1987A 
\cite{SN1987A:Detection} allowed to set upper limits on $m_\nu$.  Due
to the low statistics, the model independent bounds derived were only
at the level of $m_{\overline \nu_e}<30\,$eV \cite{Schramm:1987ra}
while more stringent limits could be obtained only under specific
assumptions \cite{SN1987A:Masses}.
More recently, a detailed reexamination of the SN1987A neutrino signal
based on a rigorous statistical analysis of the sparse data and on a
Bayesian treatment of prior informations on the SN explosion
mechanism, yielded the tighter bound $m_{\overline \nu_e}<5.7\,$eV
\cite{Loredo:2001rx}.

The first observation of neutrinos from a SN triggered in the years
following 1987 an intense research work aimed to refine the methods
for neutrino mass measurements, in view of a future explosion within our
Galaxy.  With respect to SN1987A, the time delay of neutrinos from a
Galactic SN would be reduced by a factor of a few due to the shorter
SN-earth distance. However, the neutrino flux on earth would increase
as the square of this factor and, most importantly, the large volumes
of the neutrino detectors presently in operation will yield a huge
gain in statistics.  In recent years several proposal have been put
forth to identify the best ways to measure the neutrino time-of-flight
delays, given the present experimental facilities.  Often, these
approaches rely on the identification of ``timing'' events that are
used as benchmarks for measuring the neutrino delays, as for example
the emission of gravitational waves in coincidence with the neutrino
burst \cite{Fargion:1981gg,Arnaud:2001gt}, the short duration $\nu_e$
neutronization peak that could allow to identify time smearing effects
\cite{Arnaud:2001gt}, 
the abrupt interruption of the neutrino flux due
to a further collapse of the star core into a black hole
\cite{BH}.  The more robust and less model dependent limits achievable
with these methods are at the level of $m_{\nu}\lsim 3\,$eV, as for
example in \cite{Totani:1998nf} where only the sudden steep raise of
the neutrino luminosity due to neutrinosphere shock-wave breakout is
used, without the need of relying on additional time benchmarks from
other astrophysical phenomena.
Tighter limits are obtained only under specific assumptions
for the original profiles of the SN neutrino emission or for the
astrophysical mechanisms that give rise to the benchmarks events.

In a recent paper \cite{Nardi:2003pr} we proposed a new method to extract
information on the neutrino mass from a high statistics SN neutrino
signal. The method allows to take advantage of the full statistics of the
signal, can be applied independently of particular astrophysical
assumptions about the characteristics of the neutrino emission (time evolution
of the neutrino luminosity and spectral parameters) and does not rely on
additional benchmarks events for timing the neutrinos time-of-flight delays.
The method relies on two basic assumptions: the first and most important one
is that inside the collapsing core neutrinos are kept in thermal equilibrium
by means of continuous interactions with the surrounding medium, and therefore
are emitted with a quasi-thermal spectrum.  Besides being a solid prediction
of any SN model, this picture was also confirmed by the duration of about 10
seconds of the SN1987A signal, that constitutes an evidence for efficient
neutrino trapping within the high density core.  According to this assumption,
a high statistic neutrino signal can be considered as a `self timing'
quantity, since the high energy part of the signal, that suffers only
negligible delays, could determine with a good approximation the
characteristics of the low energy tail, where the mass induced lags are much
larger. Therefore, no additional timing events are needed, and each neutrino,
according to its specific energy, provides a piece of information partly for
fixing the correct timing and partly for measuring the time delays.  The
second hypothesis is that the time scale for the variation of the
characteristics of the neutrino spectrum is much larger than the time lags
induced by a non-vanishing mass (say, much larger than 5 ms., see
(\ref{eq:delay})).  In other words, we assume that the time evolution of the
spectral parameters as inferred from the detected sample reproduces with a
good approximation the time evolution of the neutrino spectrum at the source.
Also this assumption is quite reasonable, since it is a robust prediction of
all SN simulations
\cite{Livermore,Burrows:1991kf,Totani:1997vj,Woosley:1994ux,Raffelt:2003en}
that sizable changes in the spectral parameters occur on a time scale much
larger than 5 ms.

In Ref.  \cite{Nardi:2003pr} we carried out a number of tests in order
to evaluate the sensitivity of our approach.  A typical statistics of
several thousands of neutrino events as could be detected by
Super-Kamiokande (SK) was assumed. Synthetic neutrino signals were
generated by means of a Monte Carlo (MC) code according to the numerical
results for the neutrino luminosity and average energy profiles
resulting from the simulation of the core collapse of a 20 $M_\odot$
star published by the Livermore group \cite{Woosley:1994ux}.  The
spectral shapes were taken from the dedicated study of
Janka and Hillebrandt \cite{Janka:1989aa}.  They contained a certain
amount of non-thermal distortions that are a general outcome of self
consistent simulations of SN explosions.
Finally, also the effects of neutrino oscillations in the SN mantle were
briefly analyzed in one rather conservative case (large differences
between the average energies of the different neutrino flavors, and a
sizable mixing between the neutrino spectra).  As a result, it was
shown that the method can have enough sensitivity to allow
disentangling with good confidence a neutrino mass of  1 eV from the
massless case   \cite{Nardi:2003pr,Nardi:2004ms}.

In this paper we present important improvements on the method and a more
complete set of results. We begin in sect.~2 
with a discussion of the statistical approach put forth in
\cite{Nardi:2003pr} and we show that it can be justified on a solid
theoretical basis by means of Bayesian inference reasoning.  To verify
that the quality of the results does not depend crucially on any
particular SN model, we carry out independent analysis of two
different sets of neutrino samples, as is described in sect.~3. The
first set is generated according to the same time profiles
\cite{Woosley:1994ux} and spectral shapes \cite{Janka:1989aa} used in
our previous work \cite{Nardi:2003pr}.  The second set is generated
using the alternative time profiles obtained quite recently by the
Garching group \cite{Raffelt:2003en,Buras:Private2004}. Comparison of
the results obtained with the two different sets shows that our
procedure for fitting the neutrino masses is robust with respect to
changes in the SN model.  We also refine the treatment of the effects
of neutrino oscillations in the SN mantle. The mixed spectra are
generated by using the most recent results on SN neutrino spectra
formation \cite{Buras:2002wt,Keil:2002in} that include a proper
treatment of the contributions to $\nu_{\mu,\tau}$ opacities. We do
not include earth matter effects, since they will depend on the
specific position in the sky of the SN relative to the earth, on the
specific location of each detector and on the time of the day.
However, given that even with a dedicated analysis it appears quite
challenging to identify clearly these effects \cite{Dighe:2003jg}, we
believe that this neglect is of no practical importance.  We have
identified and corrected a flaw in our MC generator that was slightly
(but artificially) enriching the number of neutrinos in the low energy
tail of the distribution.  Given that low energy neutrinos carry
important informations on the mass, the sensitivity of the method was
also slightly enhanced.  The procedure of fitting the time evolution
of the neutrino spectra is described in sect.~4. With respect to
\cite{Nardi:2003pr} we have improved both in efficiency and in
precision by adopting the $\alpha$-fit function suggested in
\cite{Raffelt:2003en,Keil:2002in}. This allows for a more simple
analytical treatment of the firsts momenta of the energy
distributions, and  considerably reduces the statistical fluctuations with
respect to the numerical fits based on the `pinched' Fermi-Dirac
functions used in \cite{Nardi:2003pr}.  Our results are presented in
sect.~5. We have studied the sensitivity of two classes of present and
planned detectors: the SK and Hyper-Kamiokande (HK)
\cite{Nakamura:2002aa} water \v{C}herenkov detectors that are
characterized by large statistics, and the KamLAND
\cite{Markoff:2003tg} and LENA
\cite{LENA} scintillator detectors  characterized by a lower
energy threshold, better energy resolution, but lower statistics.
The results show that the power of the method relies mainly on 
the overall amount of neutrino events. The lower energy threshold 
and better energy resolution   of scintillator detectors do not 
compensate for the  lower statistics. 

The claim that with the detectors presently in operation the method is
sensitive to neutrino masses at the 1 eV level
\cite{Nardi:2003pr,Nardi:2004ms,Nardi:2004uz} is confirmed by the results of
the present more complete analysis.  Note that this sensitivity is seizable
better than present results from tritium $\beta$-decay experiments
\cite{Tritium}, is competitive with the most conservative limits from
neutrinoless double $\beta$-decay \cite{Neutrinoless,Bilenky:2002aw}, and is
less precise but much less dependent from prior assumptions than
cosmological measurements \cite{Cosmology}.  A future megaton water
\v{C}erencov detector as HK will allow for about a factor of
two improvement in the sensitivity.  However, it will not be competitive
with the next generation of tritium $\beta$-decay \cite{FutureTritium}
and neutrinoless double $\beta$-decay experiments (see \cite{Cremonesi:2002is}
and references therein).  We can conclude that the occurrence of a Galactic SN
explosion within the next few years might still provide valuable informations
on neutrino masses. However, as is briefly discussed at the end of
sect.~5,  
even in the idealized situation in which the time profiles of the SN neutrino
signal are assumed known a priori, the sensitivity of these measurements
remains approximately at the level $\sim 1\,$eV (at SK).  Therefore, as new
laboratory experiments and cosmic observations will push the neutrino mass
limits sensibly below 1$\,$eV, the corresponding effects of the neutrino time
of flight delays on a SN signal will become unmeasurable.


\smallskip
\section{Outline of the statistical method} 
\label{sec:statistics}

In real time detectors, supernova electron antineutrinos are revealed
through to the positrons they produce via charged current
interactions, that provide good energy informations as well.  Each
$\bar\nu_e$ event corresponds to a pair of energy and time
measurements $ (E_i, t_i)$ together with their associated errors.  In
order to extract the maximum of information from a high statistics SN
neutrino signal, all the neutrino events have to be used in
constructing a suitable statistical distribution, as for example the
Likelihood function, that can be schematically written as
\beq
\lik \equiv  \prod_{i} {\cal L}_i = 
    \prod_{i}\>\big\{
\phi(t_i) \times F(E_i;t_i)\times \sigma(E_i)\big\}\,. 
\label{eq:schematic}
\eeq
${\cal L}_i$ represents the contribution to the Likelihood of a single
event, with the index $i$ running over the entire set of events,
$\sigma(E)$ is the $\bar\nu_e$ detection cross-section which is a well
known function of the neutrino energy
\cite{Strumia:2003zx,Vogel:1999zy} while $F(E;t)$ is the energy
spectrum of the neutrinos whose time profile is determined by the time
evolution of some suitable spectral parameters.  According to the
first assumption in the previous section, the spectrum can be
reasonably described by a quasi-thermal (analytical) distribution. If
for example a distorted Fermi-Dirac function is used, as was done in
\cite{Nardi:2003pr}, $F(E;t)$ can be parametrized in terms of a time
dependent effective temperature and a `pinching' factor
\cite{Janka:1989aa} describing the spectral distortions, and according
to the second assumption, the time dependence of the relevant spectral
parameters can be inferred directly from the data.  Therefore, the
main problem in constructing the Likelihood (\ref{eq:schematic}) is
represented by the first factor $\phi(t)$, that is the time profile of
the neutrino flux.  The strategy outlined in \cite{Nardi:2003pr} was
to find a suitable class of parametric analytical functions that could
fit reasonably well the {\it detected} flux. Given that the time
delays of the neutrinos of lowest energy are still only of the order
of a few milliseconds, it seems reasonable to assume that the same
parametric functions could also fit well the flux profile at the
source. In addition, the fact that the induced delays have a very
simple dependence on the neutrino energy and affect the signal in the
same way, independently of the specific time of the neutrino emission,
yielded us to expect that maximizing the Likelihood would allow to pin
down in an independent way the best-fit flux parameters and the
neutrino mass.  Confidence regions for the neutrino masses were found
by marginalizing $\lik$ with respect to the flux parameters, and at
each step of our analysis a special care was put in checking that no
large correlations between the flux parameters and the mass would be
present. This was interpreted as an indication of the independence of
the fitted masses not only from the flux parameters, but also from the
specific analytical profile chosen for the flux.

This procedure, that in ref. \cite{Nardi:2003pr} was justified mainly
on the basis of intuitive arguments, can in fact be put on a more
solid basis by means of Bayesian reasoning, according to which the
Likelihood function is precisely the probability of the data given
some hypothesis for their origin.  This allows us to give a well
defined statistical role to the flux profiles $\phi(t)$. Moreover, the
marginalization procedure followed in \cite{Nardi:2003pr} can be put
in direct relation with the integration of nuisance parameters
specific of Bayesian methods.
In the remaining part of this section we give a brief introduction to the
main concepts of Bayesian inference that we will use.  A short, self
contained and physics oriented introduction to Bayesian statistics
can be found in \cite{Loredo:2001rx}, while a more complete review of
Bayesian techniques and their applications in physics data analysis is
given in \cite{DAgostini:2003qr}.
 
In Bayesian inference, the degree of credibility that is assigned to a
model on the basis of certain empirical evidence, must be weighted
according to the previous knowledge of the problem (the prior).  The
central logical proposition at the basis of Bayesian statistics is
Bayes theorem:
\beq
p(M|D,I) = p(D|M,I) \times p(M|I)/p(D|I)\,.
\label{eq:bayestheo}
\eeq
The meaning of the notation $p(x|y)$ is the probability of proposition
$x$ given that $y$ is true.  The probability $p(M|D,I)$ is called the
{\it posterior probability} for model $M$ given the data $D$ and some
background information $I$; $p(D|M,I)$ is the probability that the
data $D$ are described by model $M$ and it is called the {\it sampling
probability} for $D$ or the {\it Likelihood} for model $M$; $p(M|I)$
is the {\it prior probability} for model $M$ in the absence of $D$,
and $p(D|I)$ is called the {\it evidence} for $D$ and represents the
probability that the measurement produce the data $D$ for the entire
class of hypotheses.  When $M$ is described by a (continuous) set of
parameters collectively denoted as $\Lambda $, the posterior
probability $p(\Lambda |D,I)$ becomes a multivariate probability
distribution function (pdf) for the parameters, while the Likelihood
$p(D|\Lambda ,I)$, that we will  denote by the symbol ${\cal
L}(D;\Lambda)$ in spite of its explicit dependence is not by itself a
pdf for the parameters.  The evidence $p(D|I)$ is independent of
$\Lambda$ and plays simply the role of the pdf normalization constant
$N\equiv p(D|I)=\int d\Lambda\, {\cal L}(D;\Lambda)\, p(\Lambda|I) $.

Often one is interested just in a subset of the parameters. For
example in this work $\Lambda=(m^2_\nu,\lambda)$ and we will be interested in
the implications of the SN neutrino data for the neutrino mass square
$m^2_\nu\,$, irrespectively of the particular values of the other model
parameters $\lambda$, that therefore are called {\it nuisance parameters}.
The posterior pdf for the parameter of interest is called the {\it marginal
posterior probability}, and is given by a {\it marginalization} procedure,
namely by integrating the posterior probability with respect to the nuisance
parameters:
\beq
p(m^2_\nu|D,I)  =  \int{d\lambda\; p(m^2_\nu,\lambda|D,I)} 
            =  N^{-1}\;\int{d \lambda\; {\cal L}(D;m^2_\nu,\lambda)\; p(m^2_\nu,\lambda|I)}\,.
\label{eq:marginalization}
\eeq
In practice, as is often done, we will use flat priors for all the
model parameters $\lambda$ and a step function $\Theta(m^2_\nu)=1,\,(0)$ for 
$m^2_\nu\geq 0,\,(<0)$  to exclude unphysical values of the neutrino
mass. Therefore the neutrino mass square pdf, given the SN neutrino
data $D$, reads 
\beq
p(m^2_\nu|D,I)  =  \Theta(m^2_\nu)\; \int{d\lambda\;  {\cal L}(D;m^2_\nu,\lambda)}
\label{eq:pdf}
\eeq
where the normalization constant has been absorbed for simplicity in the
Likelihood function.  The posterior pdf (\ref{eq:pdf}) is what we will use in
sect.~5 to estimate credible regions and upper limits for the neutrino
mass.  Note that we could have assumed a different prior for the neutrino mass
square, for example by introducing a second step function to exclude mass
values larger than the tritium $\beta$ decay upper limit \cite{Tritium}. This
is the way Bayesian inference allows one to take advantage of prior
informations on physical quantities. However, when the data under analysis are
informative, as is in our case, a change in the prior makes little difference
on the results. More subtle is the use of a flat prior in $m_\nu$ rather than
in $m^2_\nu$.  Throughout our analysis we will use $m^2_\nu$ not only to avoid
the problem of double maxima that would be encountered in maximizing $\cal L$
with respect to $m_\nu$, but also because $m^2_\nu$ is the relevant physical
parameter for computing the neutrino time lags.  Note that a flat prior in
$m_\nu$ would imply for the pdf $p(m_\nu|D,I)\sim |m_\nu|\,p(m^2_\nu|D,I)$ and
therefore it would favor credible regions located at smaller values of the
mass. However, by comparing results obtained with both types of priors, we
have verified that there is enough information in the data to make of little
difference which specific prior is used in estimating the credible regions and
the mass upper limits.

Coming back to the problem of constructing the Likelihood function,
and in particular of choosing a specific time profile for the neutrino
flux (namely the model $M$) we have proceeded according to the
following requirements: {\it i)} the analytical flux function must go
to zero at the origin and at infinity; {\it ii)} it must contain at
least two time scales, corresponding to the two main physical
processes responsible for neutrino emission from the star core: the
initial, fast rising phase of shock-wave breakout and accretion, and
the later Kelvin-Helmholtz cooling phase; {\it iii)} it must contain
the minimum possible number of free parameters to avoid degenerate
directions in parameter space. Still, it must be sufficiently `adaptive'
to fit in a satisfactory way the numerical flux profiles resulting
from different SN simulations, as well as flavor mixed profiles as
would result from neutrino oscillations (see  
sect.~3B). 

\begin{figure}[t]
\vskip-2mm
\hskip-6mm
\epsfxsize=80mm
 \epsfbox{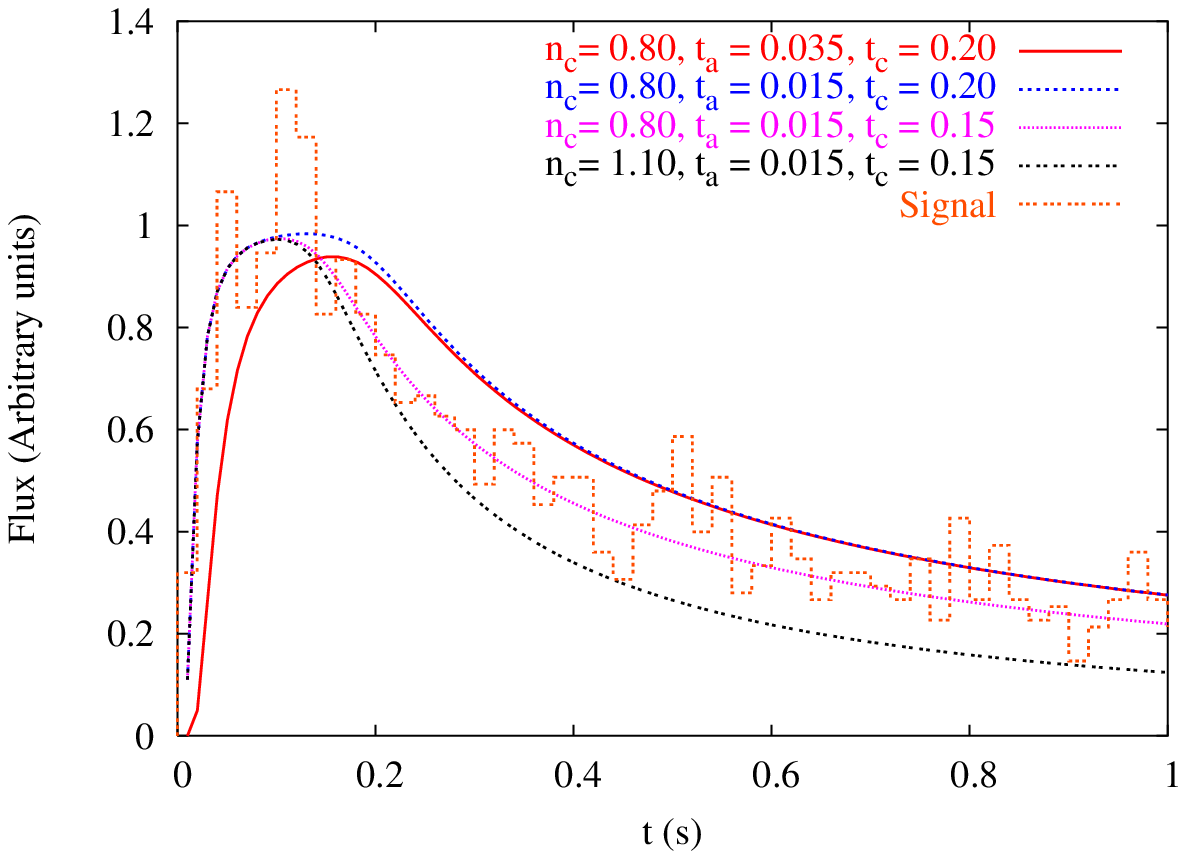}
\epsfxsize=80mm
 \epsfbox{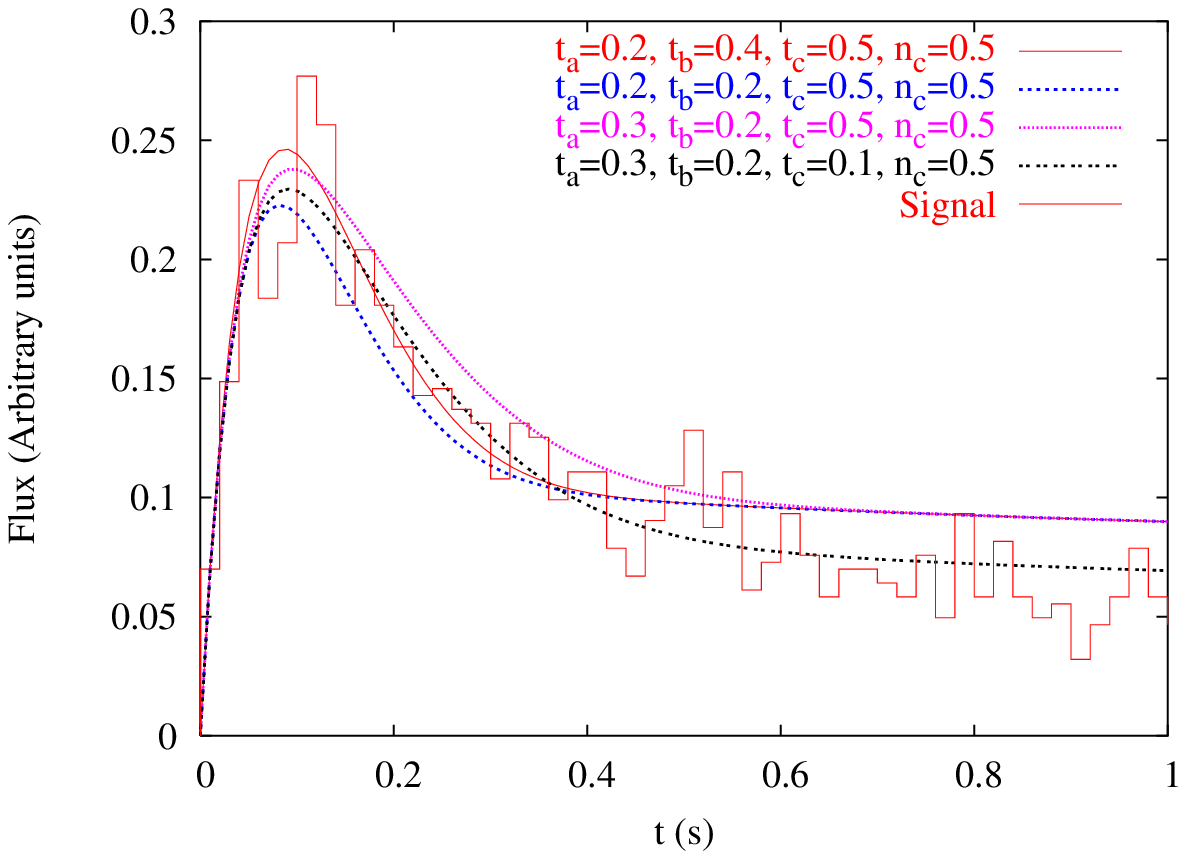}
\vskip-4mm
\caption[]{\baselineskip 12pt
  The two flux profiles discussed in sect.~2.
  In both panels, the histogram corresponds to a time binning of a 
  SN signal generated according to the Garching group simulation (SN model 2) 
  \cite{Raffelt:2003en,Buras:Private2004}. The left panel depicts the
  analytical time-profile (\ref{eq:fluxmodel1}) that has been used in our
  analysis for a few different choices of the relevant parameters ($n_a=2$ and
  $n_p=8$ have been held fixed for simplicity).  The right panel shows the
  alternative flux profile (\ref{eq:fluxmodel2}) for $n_a=n_b=1$, $A=2$,
  $C=0.8$, and a few different choices of the other parameters.
}
\label{fig:flux-models}
\end{figure}

The following model for the flux, in spite of being very simple, has
all the required behaviors, and moreover it showed a remarkable level
of smoothness and stability with respect to numerical extremization and
multi-parameter integrations:
\beq
\nonumber
\phi(t;\lambda) = \frac{ e^{-(t_a/t)^{n_a}}}{[1 + (t/t_c)^{n_p}]^{n_c/n_p}}
\casos{ll}
\sim e^{-(t_a/t)^{n_a}} & (t \to 0) \\
\sim (t_c/t)^{n_c} & (t \to \infty)\,,
\fincasos
\label{eq:fluxmodel1}
\eeq
where an overall normalization factor has been omitted for simplicity.  This
model has five free parameters that on the l.h.s of (\ref{eq:fluxmodel1}) have
been collectively denoted with $\lambda$: two time scales $t_a$ for the
initial exponentially fast rising phase and $t_c$ for the power law cooling
phase, two exponents $n_a$ and $n_c$ that control the detailed rates for these
two phases, and one additional exponent $n_p$ that mainly determines the width
of the ``plateau'' between the two phases (see fig.~\ref{fig:flux-models}).
Given that in the Likelihood analysis we will set the origin of times in
coincidence with the first neutrino detected and 
this obviously cannot correspond to the origin of time of the flux function 
(\ref{eq:fluxmodel1}) since $\phi(0,\lambda)=0$, a sixth parameter $\delta t$ is needed to allow the
function to shift freely along the time axis according to $\phi(t)\to
\phi(t+\delta t)$.  Note that the function in (\ref{eq:fluxmodel1}) is nothing
else that a physically more transparent re-parametrization of the flux model
first introduced in \cite{Nardi:2003pr}.

How much our results on the neutrino mass will depend on the specific flux
profile that has been chosen?  To answer this question we have carried out a
set of tests by using another flux model probably better motivated on
astrophysical grounds, and that was thoroughly studied in
\cite{Loredo:2001rx}:
\beq
\tilde\phi(t;\lambda) \sim 
\frac{A\,  e^{-(t/t_{a})^{n_{a}}}}{ (1+t/t_b)^{n_b}} +
  \frac{C}{(1+t/t_c)^{n_c}}\,. 
\label{eq:fluxmodel2}
\eeq
This profile is constructed by combining a truncated accretion
component (first term) with a power law cooling component (second
term). In the analysis of \cite{Loredo:2001rx} this kind of models
proved to give the best fits to the SN1987A neutrino data.  To enforce
the correct behavior $\tilde\phi(t) \to 0$ for $t \to 0$ we have
multiplied (\ref{eq:fluxmodel2}) by a suitable exponential factor.
The profiles of the two flux models are depicted in
fig.~\ref{fig:flux-models} for a few different choices of the relevant
parameters, and compared with a typical flux histogram from our MC
generator. For the case shown in the figure, the neutrino sample was
generated according to the results of the SN simulation given in
\cite{Raffelt:2003en,Buras:Private2004} (see sect.~3).
We have carried out a set of statistical tests with a few synthetic
neutrino samples using our flux model (\ref{eq:fluxmodel1}) and the
more complicated profile (\ref{eq:fluxmodel2}). Within statistical
fluctuations, the results for the neutrino mass best fits, credible
regions and upper limits, showed a high degree of consistency.  Again,
this is a firm indication that the SN data are indeed informative on
neutrino mass values of the order of 1 eV, and that our procedure is
robust not only with respect to changes in specific priors, but also
with respect to different choices of the analytical time
profiles  for  the neutrino flux.

%

\smallskip
\section{Generation of the Supernova Neutrino Signals} 
\label{sec:supernova}

The last decades have witnessed a continuous and intense effort in the
development and improvement of numerical simulations of the core collapse of
massive stars.  In spite of the important achievements in the theoretical
understanding of the underlying explosion mechanism and of the huge gain in
processing speed of modern computers, it is still unclear if the set of
physical inputs of present SN simulations is able to produce successful
explosions, and it might well be that some clue ingredients to the whole
collapse/explosion process is still missing \cite{Buras:2003sn}.  Clearly,
this somewhat weakens our confidence about the reliability of the detailed
results from the numerical simulations and, specifically for our study, about
the average energy and flux time profiles of the neutrino emission.  In
particular, different simulations produce quite diverse patterns for the time
evolution of the average energy of the different neutrino flavors, and also
the approximate values of the ratios between the amounts of energy carried
away by $e$, $\mu$ and $\tau$ (anti)neutrinos remains an issue still under
debate \cite{Keil:2002in,Raffelt:2003en}.  These two points acquire special
importance in view of the recent experimental evidences for neutrino
oscillations, that imply that the SN $\bar \nu_e$ energy spectrum that we will
observe on earth will most likely correspond to a superposition of the spectra
of different flavors.

\begin{figure}[t]
  \vskip-2mm \hskip-6mm
\epsfxsize=80mm
 \epsfbox{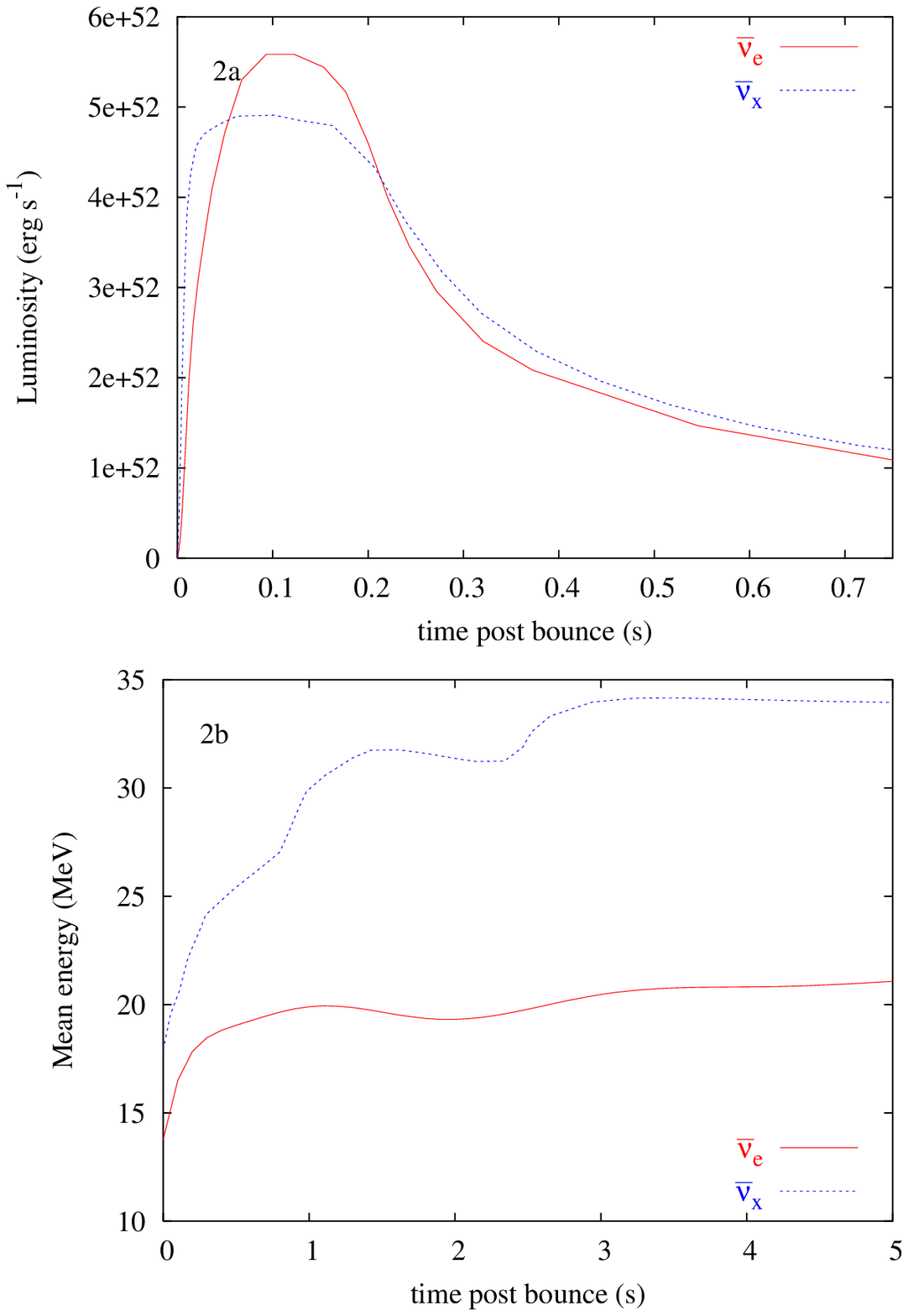}
\epsfxsize=80mm
 \epsfbox{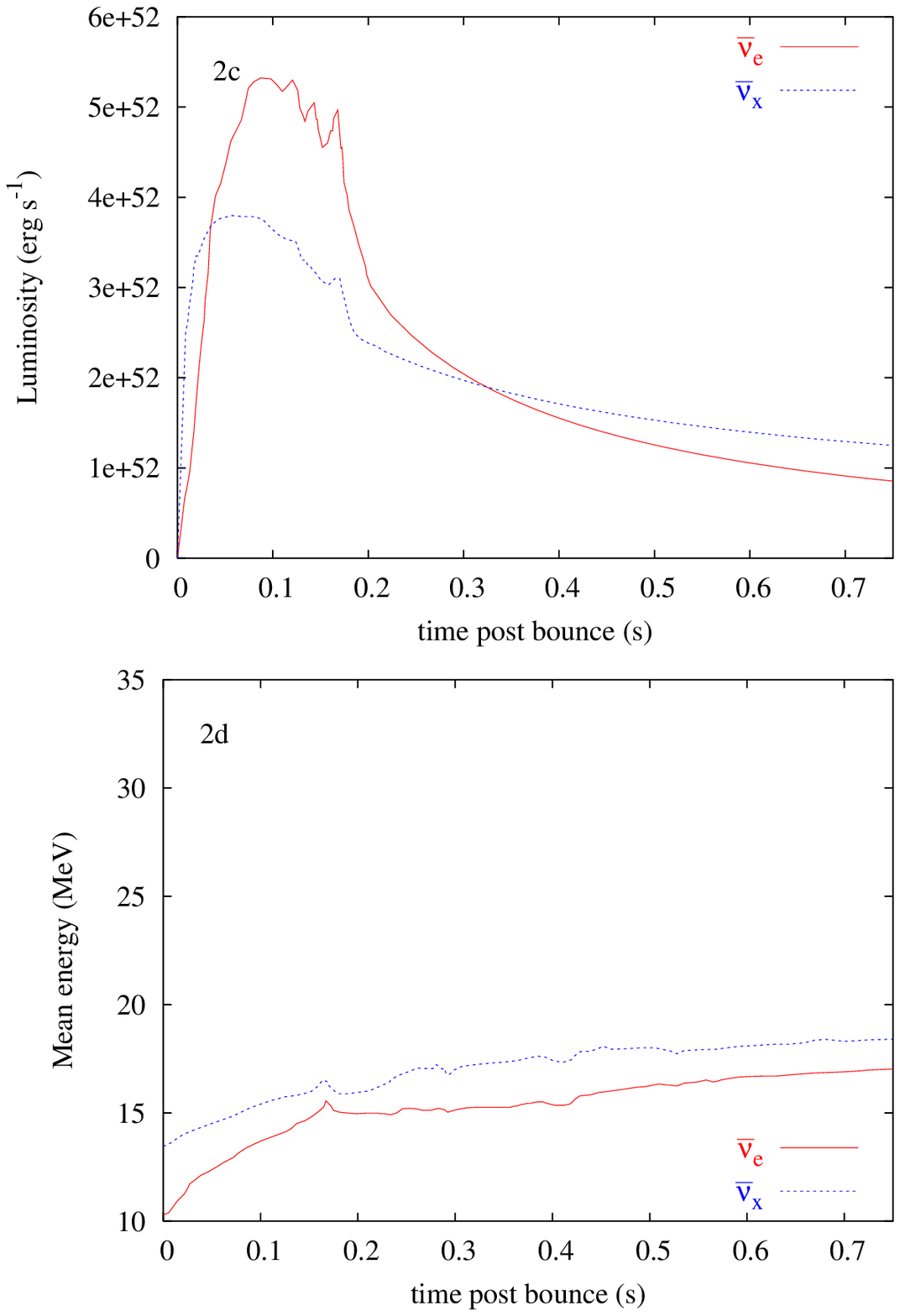}
\vskip-4mm
\caption[]{\baselineskip 12pt
  The $\bar\nu_e$ (solid lines) and $\bar\nu_{\mu,\tau}$ (dotted
 lines) time profiles for the luminosities and mean energies for the
 two different SN models described in the text. Left panels correspond
 to model 1 \cite{Woosley:1994ux} and right panels correspond to model
 2 \cite{Raffelt:2003en,Buras:Private2004}. To show how in model 1 the
 spectral differences between different neutrino flavors is increasing
 at later times, the time axes in panel 2b has been extended up to 5
 sec.
}
\label{fig:models-supernova}
\end{figure}

In order to estimate to what extent the conclusions of our study could depend
on the specific results of a given SN simulation, we have applied the method
to two different SN models, that are characterized by neutrino spectra that
fall close to the two extremes of the allowed range of possibilities.  The
first SN model, which was also used in our previous analysis in
\cite{Nardi:2003pr} and that we will denote as {\it Supernova model 1},
corresponds to a simulation of the core collapse of a 20 $M_\odot$ star
\cite{Woosley:1994ux} that was carried out with the Livermore group code
\cite{Livermore}.  The neutrino time profiles resulting from this simulation
are depicted in the left panels of fig.~\ref{fig:models-supernova}.  The
electron and $\mu\,,\tau$ antineutrino fluxes are shown in the left-upper
panel (fig.~\ref{fig:models-supernova}$a$) while the time evolution of the
neutrinos average energy is shown in the left-lower panel
(fig.~\ref{fig:models-supernova}$b$).  According to
\cite{Buras:2002wt,Keil:2002in,Raffelt:2003en}, in this simulation (as well as
in other simulations previously published) the $\mu$ and $\tau$ (anti)neutrino
opacities were treated in a simplified way. This is because these flavors are
less important than the electron (anti)neutrinos for determining the core
evolution and the SN explosion. The lack of inclusion of important
contributions to the opacities is responsible for large (and probably
unrealistic) differences in the $\bar \nu_{\mu,\tau}$ average energies with
respect to $\bar\nu_e\,$, and also results into approximate equipartition of
the emitted total energy between the six neutrino flavors.  Since the
simplified treatment of $\mu$ and $\tau$ (anti)neutrino opacities has been a
common approximation adopted in the past by several groups, large neutrino
spectral differences (up to a factor of two, see
fig.~\ref{fig:models-supernova}$b$) together with approximate energy
equipartition was established as the standard picture for SN neutrino
emission.  The second model, that will be denoted as {\it Supernova model 2},
corresponds to a recent state-of-the-art hydrodynamic simulation of a 15
$M_\odot$ progenitor star \cite{Raffelt:2003en,Buras:Private2004} carried out
by means of the Garching group code \cite{GarchingCode}.  This simulation
includes a more complete treatment of neutrino opacities
\cite{Buras:2002wt,Keil:2002in,Raffelt:2003en} and results in a quite
different picture for the neutrino spectral properties and energy repartition:
the spectra of antineutrinos of the different flavors do not differ for more
than about 20\% (fig.~\ref{fig:models-supernova}$d$) while flavor energy
equipartition appears to be violated by large factors
\cite{Buras:2002wt,Keil:2002in,Raffelt:2003en}.

%

The starting point for studying what informations on neutrino masses could be
extracted from a measurement of SN neutrinos is to generate by means of a MC
a set of synthetic measurements that hopefully will resemble closely
the results of real measurements.  This is  achieved with three main steps:
firstly, we have to generate different signals for the different neutrino
flavors as they are produced at the source; next, we have to take into account
the effects of oscillations in the SN mantle that will mix different fluxes
and spectra (as already stated, we neglect earth matter effects); finally the
specific characteristics of the different detectors (fiducial volumes, energy
thresholds and resolutions) have to be properly accounted for.  We will now
give a brief description of each one of these steps.


 \smallskip
 \subsection{Neutrino fluxes and spectra at the source} 
 \label{subsec:signal}

\begin{figure}[t]
  \vskip-2mm \hskip-6mm
\epsfxsize=80mm
 \epsfbox{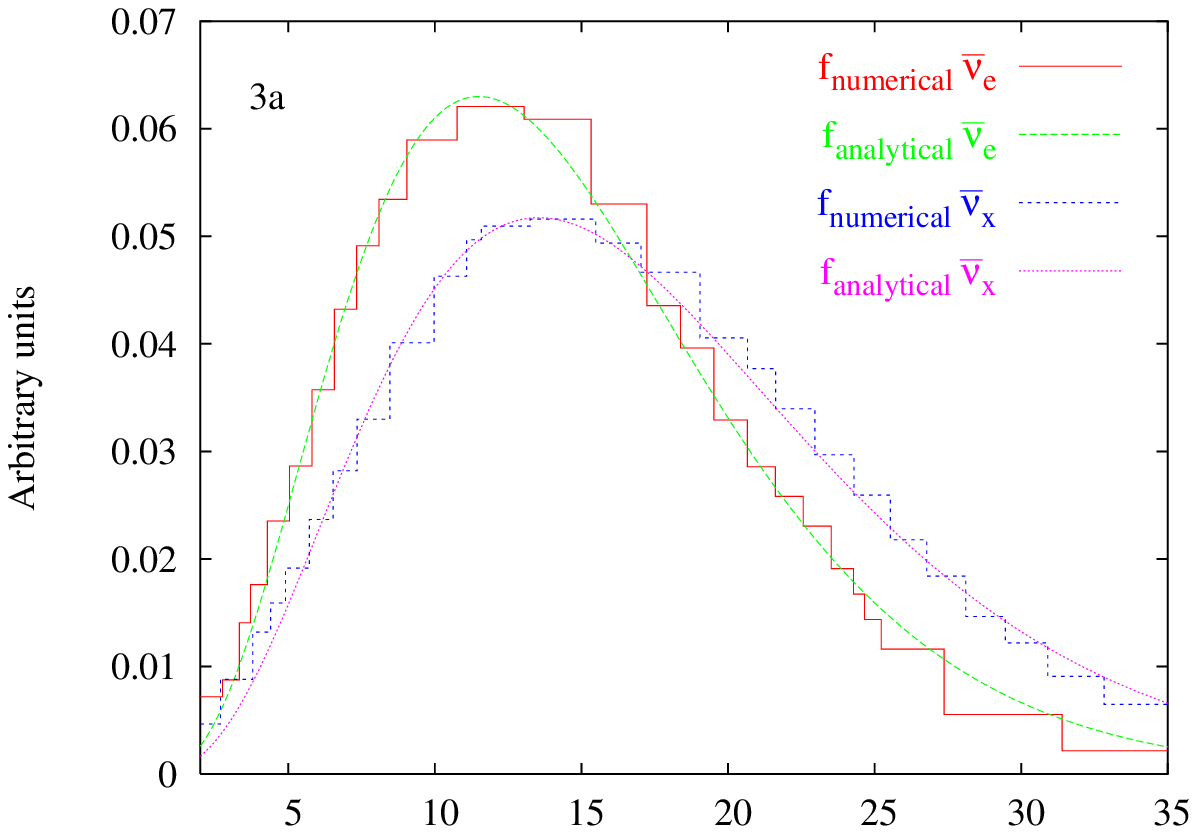}
\epsfxsize=80mm
  \epsfbox{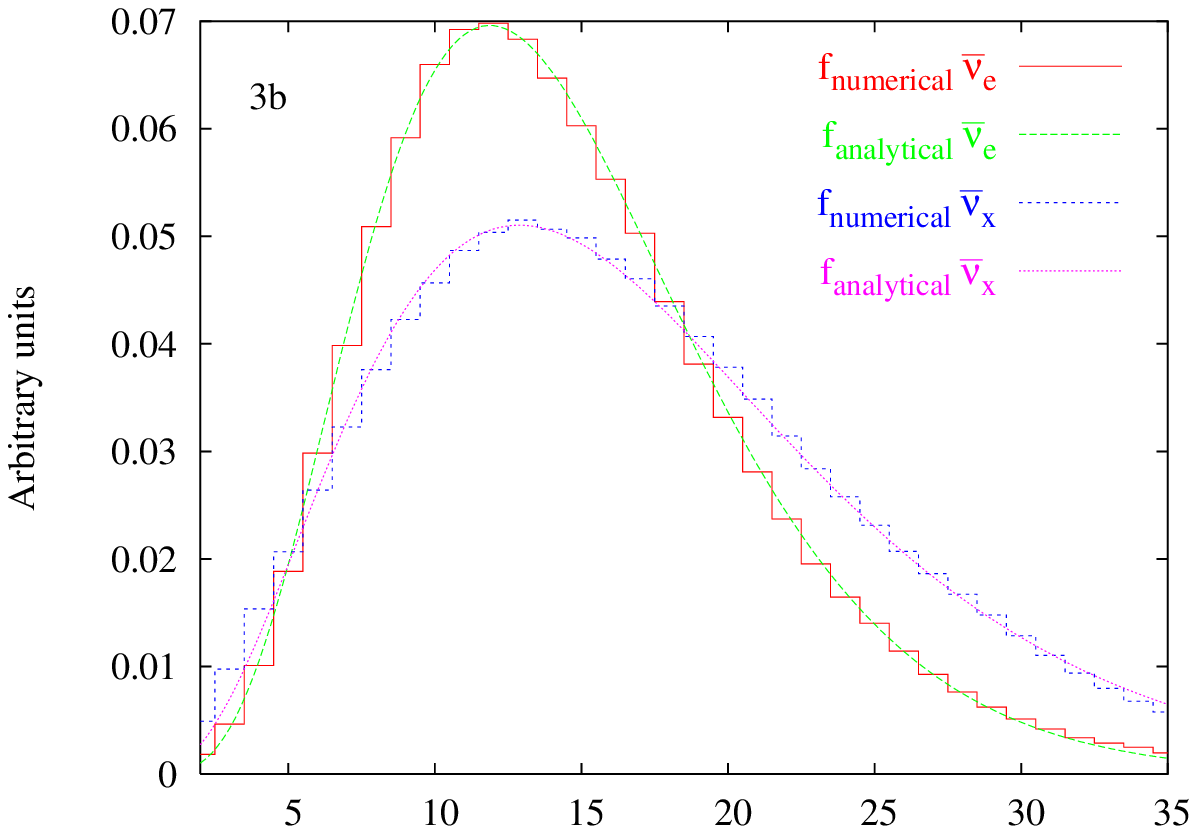}
\vskip-4mm
\caption[]{\baselineskip 12pt
  Snapshots of the neutrino spectra for the different flavors at $100$ ms
  after core bounce. Fig.~\ref{fig:numerical-spectra}a: SN model 1 (adapted
  from \cite{Janka:1989aa}). Fig.~\ref{fig:numerical-spectra}b: SN model 2
  \cite{Buras:Private2004,Janka:2004tt}.
}
\label{fig:numerical-spectra}
\end{figure}

In order to carry out a proper treatment of the emission and propagation of
the neutrino to the earth, including the effects of oscillations, we need to
know the time and energy dependence of the neutrino signal $S_\a(E,t)$ at the
emission point for each flavor $\alpha$:
\beq
S_\a(E,t) = \phi^{\rm em}_\a(t)\> F^{\rm em}_\alpha(E;t)\,, 
\qquad\qquad 
\phi^{\rm em}_\a(t) = \frac{L_\a(t)}{\barr{E}_\a(t)}\,, 
 \label{eq:emittedflux}
\eeq
where $L_\a(t)$ is the luminosity, ${\barr{E}_\a(t)}$ is the average
energy, and $F^{\rm em}_\alpha(E;t)$ is the original energy spectrum
for $\bar\nu_\alpha$.  Both SN models 1 and 2 do not provide the
complete set of informations needed to generate our samples (we
generate signals of the duration of $20\,$sec).  The results of model
1 include neutrino luminosities $L_\a(t)$ and average energies time
profiles ${\barr{E}_\a(t)}$ of the required duration.  However, the
detailed spectral shapes $F^{\rm em}_\alpha(E;t)$ are not
given~\cite{Woosley:1994ux}. To obviate this we have adopted the
numerical spectra from the detailed study presented in
\cite{Janka:1989aa}. Snapshots of these spectra taken at $100$
ms. after bounce are reproduced in fig.~\ref{fig:numerical-spectra}a.
At each instant $t$ we rescale the spectra so that the evolution of
the average energy ${\barr{E}_\a(t)}$ is correctly matched. For SN
model 2 we have used the luminosities, average energies and second
momenta of the energy distributions directly from the original
simulation \cite{Raffelt:2003en,Buras:Private2004}.  However, this
simulation was stopped after 750 ms., and the results were not
completely reliable already after the firsts 300
ms. \cite{Buras:Private2004}.  Therefore, we had to extrapolate the
results to later times.  For the luminosities we have assumed a power
law decay in agreement with general results of SN simulations
\cite{Livermore,Burrows:1991kf,Totani:1997vj,Woosley:1994ux,Raffelt:2003en}
while for the mean energies we have assumed a mild decrease after 750
ms.

%

\smallskip
\subsection{Supernova Neutrino oscillations} 
\label{subsec:oscillations}

On their way out from the high density core to the outer low density regions
of the SN mantle neutrinos will undergo flavor oscillations. Neutrino
conversion will mainly occur in crossing resonant layers where the difference
between the effective potentials felt by the different neutrino flavors is
close to the mass square difference between two mass eigenstates.  Two
resonant layers are important for the neutrino conversion process, the first
one is associated with the atmospheric neutrinos mass square difference
$\Delta m_{\oplus}^2 \simeq 2.2\times 10^{-3}\,$eV$^2$ \cite{Maltoni:2004ei}
and the second one with the solar neutrinos mass square difference $ \Delta
m_{\odot}^2\simeq 8.2 \times 10^{-5}\,$eV$^2$
\cite{Maltoni:2004ei,Bahcall:2004ut}. As a result the flux of each neutrino
flavor as observed on earth will be and admixture of the different fluxes at
the source.  In terms of the emitted $\bar\nu_e$ and $\bar\nu_{\mu,\tau}$
signals $S_{\barr{e}}$ and $S_{\barr{x}}$ 
the $\bar\nu_e$ signal at the detector
can be written as
\beq 
L^2\, S^{\rm det}_{\barr{e}} =
p_{\barr{e}}\, S_{\barr{e}} +
\left(1-p_{\barr{e}}\right)\, S_{\barr{x}}\,, 
\label{eq:generalflux}
\eeq
where $L$ is the SN-earth distance and $p_{\barr{e}}$ is the $\bar\nu_e$
survival probability.  Note that while three different mass eigenstates
propagate incoherently from the SN to the earth and concur to determine
$S^{\rm det}_{\barr{e}}$, the mass differences are much smaller than the
sensitivity to the absolute value of the neutrino mass, and therefore
neutrinos can be treated as degenerate for all practical purposes.  We see
from (\ref{eq:generalflux}) that the observed flux can be written in terms of
just one survival probability $p_{\barr{e}}$.  This is because of two reasons:
firstly, the large hierarchy between $ \Delta m_{\oplus}^2 $ and $\Delta
m_{\odot}^2$ implies that the two resonant layers are well separated, and
therefore the conversion process can be factorized into a two flavor problem
at each layer; secondly, the $\bar\nu_\mu$ and $\bar\nu_\tau$ fluxes are equal
(both are represented by $S_{\barr{x}}$).  A careful analysis of the level
crossings encountered by the propagating eigenstates allows to write
$p_{\barr{e}}$ in terms of the two antineutrino probabilities
$\barr{P}_\oplus$ and $\barr{P}_\odot$ for jumping to a different matter
eigenstate when traversing the resonant layers \cite{Dighe:1999bi}. We need to
distinguish two possibilities: the case of normal hierarchy (NH) when
$\bar\nu_e$ has the small admixture $\sin^2\theta_{e3}< 0.047$ ($3\sigma$)
\cite{Maltoni:2004ei,Bahcall:2004ut} in the heaviest state, and the inverted
hierarchy (IH) when the small admixture is in the lightest state. Denoting by
$|U_{ei}|$  the modulus of the electron (anti)neutrino mixing with the
$i=1,2,3$ mass eigenstate, we have:
\beqa \nonumber 
{\rm (NH):\qquad\qquad}  p_{\barr{e}} &=& |U_{e1}|^2\,(1-\barr{P}_\odot) +
|U_{e2}|^2\,\barr{P}_\odot\,,  \\
 p_{\barr{e}} &\rightarrow &  \quad |U_{e1}|^2 \approx
 \cos^2\theta_\odot\,;  \\
\nonumber
{\rm (IH):\qquad\qquad}  p_{\barr{e}} &=& |U_{e1}|^2\,(1-\barr{P}_\odot)\,\barr{P}_\oplus
+|U_{e2}|^2\,\barr{P}_\odot\,\barr{P}_\oplus
+|U_{e3}|^2\,(1-\barr{P}_\oplus)\,,  \\ 
p_{\barr{e}}   &\rightarrow&  \quad  |U_{e1}|^2\,\barr{P}_\oplus
+|U_{e3}|^2\,(1-\barr{P}_\oplus)\,;   
\label{eq:survival}
\eeqa 
where in the second and last lines the adiabatic limit $\barr{P}_\odot \to 0$
has been taken.  Adiabaticity of the transitions in the layer corresponding to
the solar neutrino mass square difference is guaranteed by the results of
global fits to solar neutrino oscillations, that established the large mixing
angle solution with $\sin^2\theta_\odot\simeq 0.29$
\cite{Maltoni:2004ei,Bahcall:2004ut}.  Note that for NH the $\bar \nu_e
\leftrightarrow \bar\nu_3$ transitions are strongly suppressed due to the
smallness of $|U_{ei}|^2$, and since there are no level crossing for
$\bar\nu_3$ this state decouples and $ p_{\barr{e}}$ does not depend on
$\barr{P}_\oplus$.  In general this is not true for the IH case.  However, for
$|U_{e3}|^2\gsim 10^{-3}$ the transition is adiabatic also in the first layer
implying $\barr{P}_\oplus \approx 0$ and we obtain $p_{\barr{e}} \approx
|U_{e3}|^2 \leq 0.047$ \cite{Maltoni:2004ei,Bahcall:2004ut}.  This corresponds
to an almost complete $\bar\nu_e \leftrightarrow \bar\nu_x$ spectral swap.
For smaller values of $|U_{e3}|^2$ the transition enters the non-adiabatic
regime and we obtain $p_{\barr{e}} \approx \barr{P}_\oplus\,\cos^2\theta_\odot
$ (in this case the survival probability also depends on the neutrino energy,
though not in a strong way).  In the following we will restrict ourself to the
NH case that corresponds to the most interesting situation, since it yields a
$\bar\nu_e$ spectrum which is an admixture of about 1/3 of the harder $\bar
\nu_x$ original spectrum.  Note that the IH case in the strongly non-adiabatic
regime $(|U_{e3}|^2 \lsim 10^{-5}\,,\,\barr{P}_\oplus \approx 1)$ would also
yield the same mixed spectrum. The IH case with adiabatic transitions in the
first layer is less interesting since the almost complete $\bar\nu_e$-$\bar
\nu_x$ spectral swap would yield a single component neutrino spectrum just
with a different effective temperature, much alike the non-oscillation case.
Obviously, oscillations effects resulting in a mixed spectrum will be more
important for large spectral differences as in SN model 1, since the fits to
the energy distributions by means of a single quasi-thermal spectral function
will yield only an approximate result.  In SN model 2, where the two spectra
do not differ too much, the main effect of oscillations would be that of a
change in the statistics of the detected signal induced by deviations from
exact energy equipartition of the original fluxes, while the fits to the
energy spectrum will not be affected much.

\smallskip
\subsection{Neutrino Detection} 
\label{subsec:detection}

The double differential rate for the SN neutrino events in specific
detector reads 
\beq 
\frac{d^2\,n^{\rm }_{\ae} (E,t)}{dE\,dt} = N_T\, \int_{E_{\rm
th}}{dE'\, S^{\rm det}_\ae(E',t)\, \s(E')\, \epsilon(E')\, {\cal
R}(E,E')},
\label{eq:rate}
\eeq
where $S^{\rm det}_\ae(E,t)$ is the incoming energy and time dependent
$\bar \nu_e$ distribution (\ref{eq:generalflux}) and $\sigma(E)$ is
the cross-section, that for water \v{C}erenchov and scintillator
detectors corresponds to the inverse $\beta$ decay process of
producing a positron via $\bar\nu_e$ capture by a proton
\cite{Strumia:2003zx,Vogel:1999zy}. All the other quantities vary according to the
specific detector considered: $N_T$ is the number of target particles
in the fiducial volume, $E_{\rm th}$ is the detection energy threshold
and $\epsilon(E)$ the detection efficiency. We assume 100\% efficiency
above threshold (that is a good approximation e.g. for SK) so that
$\epsilon(E)=\theta(E-E_{\rm th})$ with $\theta$ the unit step
function.  Finally ${\cal R}(E,E')$ is the energy resolution function
that accounts for the uncertainties in the measurement of neutrino
energies.  We approximate this function with a Gaussian distribution
with mean  equal to the measured energy $E$, and standard deviation
$\Delta E$ given by \cite{Lunardini:2001pb}
\beq 
\frac{\Delta E}{\rm MeV} = a_E \sqrt{\frac{E}{\rm MeV}} + b_E \,\frac{E}{\rm MeV}\,.
\label{eq:Euncertainty}
\eeq
The specific values of $a_E$ and $b_E$ as well as other relevant parameters
for the most important SN neutrino detectors presently in operation and for a
few proposed large volume detectors are collected in
table~\ref{tab:detectors}.  In the last column of the table we also give a
range for the total number of $\bar\nu_e$ events that a Galactic SN at a
distance of 10~kpc is expected to produce in each detector, assuming the two
SN model and the oscillation pattern discussed above, and taking into account
only charged current reactions that can provide good energy and time informations.

The sets of synthetic samples to which we have applied our procedure have been
generated with a MC code where bi-dimensional rejection in $E$ and $t$ is
applied to the function (\ref{eq:rate}) describing the neutrino event rate for
each detector considered. This yields a set of energy and time pair of values
$(E_i,t_i)$ each of which corresponds to the detection of one neutrino. To
take into account the finite energy resolution, the value of $E_i$ is
obtained from an initial MC value $E'_i$ by redrawing the energy according to
the resolution function ${\cal R}(E,E')$.  Of course, because of oscillations,
the times and energies of the final samples will correspond to a superposition
of the original $\bar\nu_e$ and $\bar\nu_{\mu,\tau}$ fluxes and spectra.

\begin{table}[t]
\begin{tabular}{>{}p{3cm}c>{\centering}p{2cm}p{2cm}cc}
\hline\hline
\multicolumn{2}{c}{Detector}
& $E_{\rm th}$ 
& $(a_E,b_E)$
&  {\small Fiducial mass}
&  $N^{\det}_\ae$ \\
 & & (MeV) & & (kton) & ($L = 10\;{\rm kpc}$) \\\hline\hline
\blankline{6}
%
\cerenkov & SK\cite{Nakahata:1998pz,Beacom:2003nk} &
5 & $(0.47, 0)$ & $32$ & 5,900 - 9,990 \\
 & (H$_2$O) & & & & \\
 & SNO\cite{Virtue:2001mz,Aharmim:2004uf} & 4 & (0.35, 0) &  &  \\
 & H$_2$O &  &  & 1.4 & 260 - 440\\
 & D$_2$O &  &  & 1.0 &  80 - 160 \\
Scintillator & KamLAND \cite{Iwamoto:2003aa} 
& 2.6 & (0, 0.075) & 1.0 & 240 - 400 \\
 &  (N12+PC+PPO) & & & & \\
\hline
\blankline{6}
%
%
\cerenkov & HK\cite{Nakamura:2002aa}
& 5 & (0.5, 0) & 540 & 100,000 - 170,000 \\
&  (H$_2$O) & & & & \\
 & UNO \cite{Jung:1999jq} 
& 5 & (0.5, 0) & 650 & 120,000 - 203,000 \\
& (H$_2$O) & & & & \\
Scintillator
& LENA \cite{LENA}
& 2.6 & (0.1, 0) & 30 & 7,500 - 12,600 \\
& (PXE) & & & & \\
\hline\hline
\end{tabular}
\caption{\baselineskip 12pt 
The relevant $\bar\nu_e$ detection parameters for some of the present and
proposed detectors. In the last column we give the expected range for the
number of charged current $\ae$ events from a Galactic SN at 10~kpc, assuming
the neutrino oscillation pattern discussed in sect~3B. The larger (smaller)
numbers correspond to SN model 1 (model 2).
}
\label{tab:detectors}
\end{table}


\smallskip
\section{Construction of the Likelihood} 
\label{sec:likelihood}
We will now describe the construction of the Likelihood that is used
as a statistical estimator for the model parameters, and in particular
for the neutrino mass.  Strictly speaking, a maximum Likelihood
analysis of the whole signal should consists in a full bi-dimensional
extremization (in time and energy) of a complete SN model, thus
including the spectrum and its time evolution. However, besides
requiring the introduction of several more parameters, this would also
introduce an unpleasant model dependence, since the spectral
characteristics and in particular their time evolution are probably
the quantities that more crucially depend on the specific SN
simulation.  However, in the limit of large statistics and under the
second assumption discussed in sect.~1, the problem can be greatly
simplified by performing first, as an independent step, a fit to
the neutrino spectrum.  Namely, the time dependent spectral function
for the model can be inferred directly from the data (and therefore
without introducing any crucial model dependence) and next the result
can be input in the Likelihood analysis as a given information.
Strictly speaking, because of the statistical fluctuations affecting the
results of the spectral fit, at each new run we will be testing a
different SN model (the same flux function, but slightly different
spectral characteristics).  Nevertheless, if the statistics is large,
the models will not differ too much, and as we will see `factorizing'
the problem in this way indeed yields consistent results.

As was discussed in sect.~2,
three different terms enter the expression for the Likelihood
(\ref{eq:schematic}): the $\bar\nu_e$ detection cross section, the time
dependent spectral function and the neutrino flux time profile.  For the
cross-section we use the convenient parametrization given in
\cite{Strumia:2003zx}:
\beq
\frac{\sigma_{\bar\nu}(\bar\nu_e p \to e^+n)}{ 10^{-43}{\rm cm}^2} =
p_e E_e\, E_{\bar\nu}^{-0.07056+0.02018\ln
  E_{\bar\nu}-0.001953\ln^3E_{\bar\nu} }, 
\label{eq:cross-section}
\eeq
where $E_e=E_{\bar\nu}-\Delta_{np}$ with $ \Delta_{np}=m_n-m_p\approx
1.293\,$MeV and all the energies are in MeV.  This expression does not
take into account the effects related to the non isotropic angular
distribution of the differential cross section, discussed in detail in
\cite{Vogel:1999zy}. However, since for the relevant range of SN 
neutrinos energies the corresponding error induced on the energies of
the positrons remain safely below the experimental error, for the
present scopes eq. (\ref{eq:cross-section}) is sufficiently accurate.
We model the time dependent spectral function $F(E;t)$ by means of the
$\alpha$-distribution introduced in \cite{Raffelt:2003en,Keil:2002in}:
\beqa \nonumber
&&F\left(E,\bar\epsilon(t),\alpha(t)\right) = N(\bar\epsilon,\alpha)\> 
\left({E}/{\bar\epsilon}\right)^{\alpha}e^{-(\alpha+1)\,E/\bar\epsilon}\,,
\\
&& N(\bar\epsilon,\alpha)= 
{(\alpha+1)^{\alpha+1}}/{\Gamma(\alpha+1) \bar\epsilon}\,. 
\label{eq:alphafit}
\eeqa
Using the well known relation
$\alpha\,\Gamma(\alpha)=\Gamma(\alpha+1)$ it is easy to verify that
the function (\ref{eq:alphafit}) has the nice property of allowing a
simple analytical estimation of the two spectral parameters
$\bar\epsilon$ and $\alpha$ directly in terms of the first and second
momentum of the energy distribution:
\beq
\bar\epsilon = \langle E \rangle\,; \qquad \qquad 
\frac{2+\alpha}{1+\alpha}= \frac{\langle E^2 \rangle}{\langle E \rangle^2}\,.
\label{eq:momenta}
\eeq
Often in the literature the SN neutrino spectrum is approximated in terms of a
nominal Fermi-Dirac distribution~$\sim [1+\exp(E/T-\mu)]^{-1}$ where $T$ is an
effective temperature and $\mu$, that enters the distribution similarly to a
chemical potential, describes the spectral distortions, and similarly to
$\alpha$ in (\ref{eq:momenta}) is related to the ratio between the second and
the first energy momentum square. Such a choice was adopted in
\cite{Nardi:2003pr}, and indeed is physically well motivated since a thermal
spectrum would follow this behavior. However, starting from a discrete sample
of neutrinos, a nominal Fermi-Dirac spectrum can be reconstructed only by
carrying out numerical fits to the energy momenta until the correct values of
$T$ and $\mu$ are determined through a minimization procedure. In contrast,
the $\alpha$ distribution can be straightforwardly determined through
eqs.~(\ref{eq:momenta}). At the same time, as it was shown in
\cite{Keil:2002in}, within an energy range sufficiently large for all
practical purposes the $\alpha$ distribution is equivalent to a nominal
Fermi-Dirac to better than 10\%.  Clearly, when estimating $\bar\epsilon$ and
$\alpha$ from a set of {\it measured} neutrino energies, the effect of the
detection cross-section (\ref{eq:cross-section}) that modifies the observed
energy distribution has to be taken into account. Thus, the first and second
momentum on the r.h.s in (\ref{eq:momenta}) are computed as
\beq 
\mean{E^n} = \frac{\sum_i E_i^n/\sigma_{\bar\nu}(E_i)}
{\sum_i 1/\sigma_{\bar\nu}(E_i)}\,, \quad\qquad n=1,2\,, 
\label{eq:correctedmomenta}
\eeq
where the sum runs over all the  neutrinos belonging to
the same time window.  In order to obtain two
continuous function of time $\bar\epsilon(t)$ and $\alpha(t)$ 
eq.~(\ref{eq:correctedmomenta}) is applied to a set of windows centered in $t$
and of width $\Delta t$ that, in order to reduce statistical fluctuations, is
chosen large enough to contain a sufficient number of neutrinos (a few
hundreds). The central value of each new window is determined as
$t_{n+1}=t_n+\delta t$, with $\delta t \ll \Delta t$ so that different windows
overlap, thus ensuring that the fit to the spectral parameters yields two
smooth functions.

The last ingredient to construct the Likelihood is the neutrino flux time
profile $\phi(t;\lambda)$ eq.~(\ref{eq:fluxmodel1}) that, as discussed in
sect.~2, carries the dependence on the model parameters.  Instead than
including the dependence on $m^2_\nu$ directly in the flux function by means
of a redefinition of the time variable, it is more convenient to proceed in
the following way: given a test value of the neutrino mass, the arrival time
of each neutrino is shifted according to its time delay
eq.~(\ref{eq:delay}). After doing this, the value of the Likelihood is
computed for the time-shifted sample.  However, because of the finite
resolution the {\it measured} energies that are used to evaluate the time
shifts do not correspond to the {\it true} energies that determine the real
neutrino delays.  Therefore, even when the correct value of the test mass is
used, the time-shifted neutrino sample will not correspond exactly to the
sample originally emitted.  Although completely natural (as well as
unavoidable) this behavior can produce a dangerous situation. When the energy
measurement yields a value {\it smaller} than the true energy, a neutrino
arrival time can be shifted to a negative value where the flux function
vanishes, implying that the log-Likelihood diverges. This would imply
rejecting the particular neutrino mass value for which the divergence is
produced, regardless of the fact that it could actually be close to the true
value.  To correct this problem we adopt the following procedure.  The
contribution ${\cal L}_i$ to the Likelihood (\ref{eq:schematic}) of a neutrino
event with measured energy $E_i\pm \Delta E_i$ for which, after subtracting
the delay $\delta t_i = m^2_\nu L/2E_i^2\,$, we obtain a negative value
$t_i<0$ (or a value close to the origin of the flux function $t_i \sim 0$) is
computed by convolving it with a Gaussian ${\cal G}(t;t_i,\sigma_i)$ centered
in $t_i$ and with standard deviation $\sigma_i = 2\,\delta t_i\, \Delta
E_i/E_i\;$:
\beq 
{\cal L}_i = \int{dt\; [\phi(t) \times F(E;t) \times \sigma(E)]\>
{\cal    G}(t;t_i,\sigma_i)}. 
\label{eq:regeventprob}
\eeq
Clearly this regularization of the divergent contributions to the
log-Likelihood is physically motivated by the fact that the origin of
the problem is the uncertainty in the energy measurements, that
translates into an uncertainty in the precise location in time of the
neutrino events after the energy-dependent shifts are applied.

\begin{figure}[t]
\vskip-2mm
\hskip-6mm
\epsfxsize=160mm
 \epsfbox{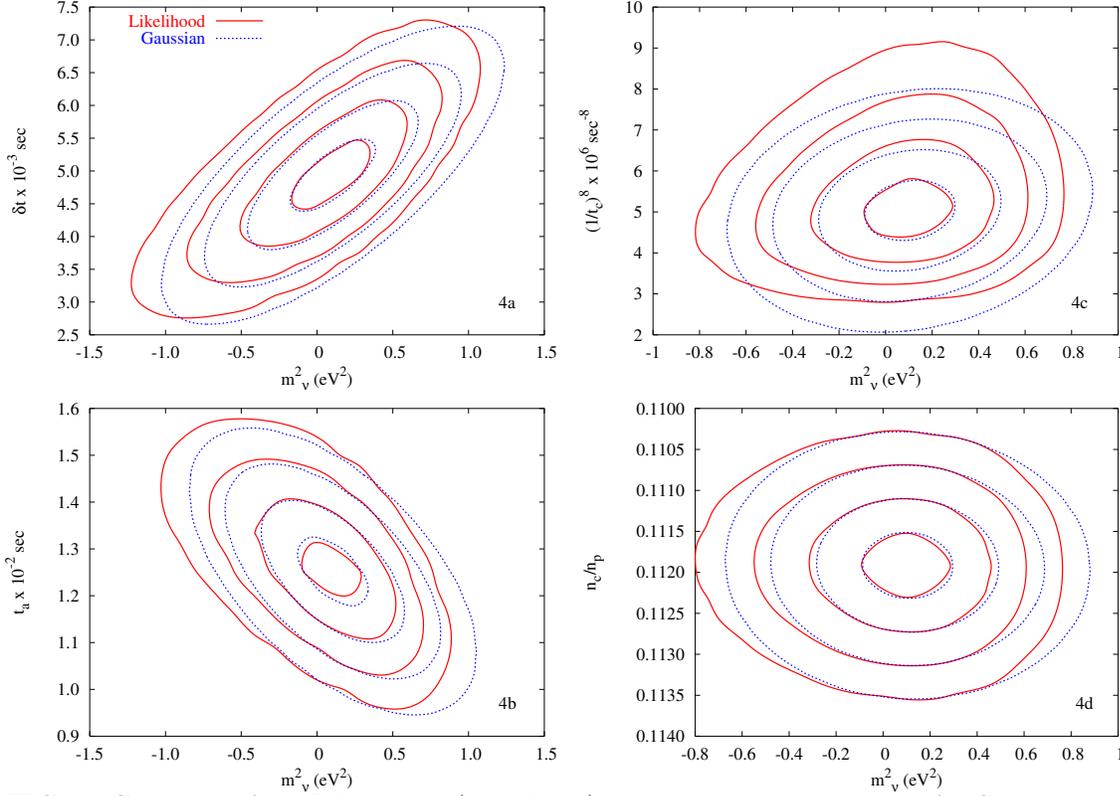}
\vskip-4mm
\caption[]{\baselineskip 12pt Contours of the Likelihood (solid lines)
  compared with contours of a Gaussian distribution (dotted lines) of the same
  mean and covariance, in four different two parameters spaces: $m^2_\nu$
  versus~\ref{fig:contours}a:~the time shift $\delta t$ of the flux function;
  \ref{fig:contours}b:~the signal raising time scale $t_a$ ($n_a=1$);
  \ref{fig:contours}c:~the time scale of the cooling phase $t_c$ ($n_c=0.8$);
  \ref{fig:contours}d:~the ratio $n_c/n_p$ (see eq.~(\ref{eq:fluxmodel1})).
  The contours correspond to $0.5,\,1.0,\,1.5$ and 2.0 $\sigma$.}
\label{fig:contours}
\end{figure} 
%

A few remarks about possible systematic errors in our procedure are in
order.  We are aware of the presence in our analysis of at least three
sources of systematics: {\it i)} the artificial stopping of the
generation of the neutrino signal at 20 sec; {\it ii)} the convolution
procedure we have just described; {\it iii)} the unfolding of the
cross section in computing the first and second momentum of the energy
distribution in eq.~(\ref{eq:momenta}).  We will give now a brief
description of each one of these effects; however, it should be
stressed that we know how they could be avoided in a real analysis and
moreover, as we will show in the next section, the overall uncertainty
in the analysis is statistically dominated and it is safe to neglect
the effects of systematic errors on the final results.

{\it i)} Strictly speaking, any procedure that interrupts the
generation or the analysis of the neutrino signal before it naturally
drops to zero can be a source of systematic errors.  To show this, let
us focus on the contributions to the log-Likelihood of neutrinos of
the same energy, say between $E$ and $E+\Delta E$ that, for a given
test mass, will all suffer the same time shift $\Delta t$ (the
generalization to the full signal case with neutrinos of all energies
is straightforward).  Let us assume that the distribution in time of
this subset of neutrinos in the original signal (that is the pdf) is
known, and let us call it $P_E(t)$. Due to the time shift, we will
have $P_E(t+\Delta t) dt$ neutrinos between $t$ and $t+d t$ that will
give a contribution $P_E(t+\Delta t)\,\log P_E(t) \,dt$ to the
log-Likelihood. Summing up the contributions of all the neutrinos up
to a finite time $t_0$, expanding in powers of $\Delta t$ and imposing
the extremization condition, we easily obtain:
 \begin{equation}\nonumber
\frac{\delta\> \log{\cal L}_E(\Delta t)}{\delta(\Delta t)}
= P_E(t_0)\> \left(\log P_E(t_0)-1\right)+ 
\sum^{+\infty}_{n=1} \frac{(\Delta t)^n}{n!} \int^{t_0}_{- \infty}
  dt\, P_E^{(n+1)}(t)\; \log P_E(t)=0\,.
\label{eq:extreme}
\end{equation}
In the limit $t_0 \to\infty $ the first term on the r.h.s vanishes
since $P_E(t_0)\to 0$ as is required for any normalizable pdf, and
therefore the extremization condition is satisfied for $\Delta
t=0$. In contrast, if $P_E(t_0)\neq 0$ then (\ref{eq:extreme}) is not
satisfied for $\Delta t=0$ and one obtains an incorrect result.  However,
if $t_0 \gg 0$ and $F(t_0)\approx 0$, as is our case in cutting the
signal at 20 sec, a good approximation to the correct answer is found,
and for this reason the systematic error induced by this effect on our
results is negligible. Of course, for a real signal the analysis will
have to be carried out up to the last neutrino detected, very likely
much beyond the 20~sec limit we have been using for convenience, and
therefore we do not have to worry for this kind of systematics.

{\it ii)} The convolution procedure described by
eq.~(\ref{eq:regeventprob}) induces a second source of systematic
errors. This is because fast and accurate minimization routines rely
on the knowledge of first derivatives, and hardly tolerate any
`jump'. Therefore when, because of the scanning of different mass
values, a neutrino event is shifted to time values for which the flux
function is not vanishing, convolution cannot be switched off
abruptly, since this can result in the abnormal termination of the
minimization routine.  Instead, convolution has to be turned off
`adiabatically', by reducing continuously the width of the convolution
region while moving toward times where the flux function starts
raising. However, the time variation of the flux is rather sharp, and
this can slightly alter the contributions to the log-Likelihood from
the early part of the signal.  In our analysis also this effect is
negligible. In the case of a real signal, robust but rather slow
non-derivative minimization routines, like MC minimization, could be
used thus avoiding the whole problem at once.

{\it iii)} To reconstruct the time evolution of the neutrino energy
spectrum the effect of the cross-section that modifies the observed
energy distribution must be accounted for. However, the expression
given in eq.~(\ref{eq:momenta}) represents only an approximation to
the exact unfolding of the cross section. This is because a neutrino
of energy $E$ is detected with probability proportional to
$\sigma_{\bar \nu}(E)$ but, because of the detector finite energy
resolution, its energy is measured to be $E+\Delta E$. Therefore, when
unfolding the cross section $\sigma_{\bar\nu}(E+\Delta E)$ is used,
since the true value of the energy is unknown.  This affects the
estimate of the momenta of the distribution by terms that are formally
of order $ \langle \dots (\Delta E)^2\rangle$ where the dots stand for
the relevant combinations of powers of $E$ and derivatives of
$\sigma_{\bar\nu}(E)$. We have verified that the overall effect of the
approximation represented by eq.~(\ref{eq:momenta}) in reconstructing
the time evolution of the energy distribution is observable but small,
and that the systematic error induced on the fits to the neutrino
masses is negligible.  Clearly, also this effect can be accounted for
in a real analysis by estimating the expectation values of the
relevant terms of order $(\Delta E)^2$ and by properly correcting for
this the inferred values of the energy momenta.


\smallskip
\section{Results and discussion} 
\label{sec:results}

Once the Likelihood is constructed according to the procedure
described in the previous section, a statistical study of the
sensitivity of the SN neutrino signal to the neutrino mass can be
carried out.  According to eq.~(\ref{eq:pdf}), the marginal posterior
pdf $p(m^2_\nu|D,I)$ is obtained by marginalizing the Likelihood with
respect to the nuisance (flux) parameters. However, the CPU time
required to carry out all the necessary multidimensional integrations
would be exceedingly large, especially considering that we need to
analyze a large set of neutrino samples, corresponding to different SN
models, SN-earth distances and also to different detectors. Therefore,
as is often done in this situation, we will approximate the marginal
posterior probability with the {\it Profile Likelihood} (PL)
$\hat{\cal L}(D|m^2_\nu)$, that corresponds to the trajectory in
parameter space along which, for each given value of $m^2_\nu$, the
Likelihood is maximized with respect to all the other parameters. It
can be shown that for a multivariate Gaussian the PL coincides with
the marginal posterior $p(m^2_\nu|D,I)$, and therefore our results
will be reliable to the extent the Likelihood approximates well enough
a normal distribution.  In fig.~\ref{fig:contours} we compare
different parameter space contours for ${\cal L}(D;m^2_\nu,\lambda)$
with those of a corresponding normal distribution constructed from the
set of second derivatives in the maximum.  We see that within the
region where the contributions to the integrations are dominant, the
Likelihood approximates rather well a Gaussian distribution.

In spite of the fact that the contours in fig.~\ref{fig:contours}
appear to be sufficiently close to the Gaussian ones to justify
the use of the Profile Likelihood, there are at least two known
effects that imply the presence in the analysis of a certain amount of
non-Gaussian features, and some care should be put in deriving
numerical results.

{\it i)} Even if each distribution is approximately Gaussian for a wide
range of $m^2_\nu$, there is always a value of of the neutrino mass
square for which the distribution is cut to zero.  To give an example,
in a standard Likelihood analysis the detection of just one neutrino
of 7~MeV from a SN at 10~kpc, 10~ms after the onset of the signal
would by itself be sufficient to exclude a neutrino mass of 1 eV.
This is because in evaluating the Likelihood for a test mass
$\gsim 1\,$eV the contribution of this neutrino would vanish (due to
$\phi(t)\to 0$) driving to zero the whole Likelihood.  If the error on
the energy measurement is taken into account, see
eq. (\ref{eq:regeventprob}), this effect is smeared but its
non-Gaussian nature is not changed.  Therefore, strictly speaking,
inferring a limit at some c.l. from the width of the distribution
(say, from the second derivative with respect to $m^2_\nu$ in the
maximum) would only yield an upper bound on the limit, but not the
true limit.  Reliable limits can be obtained only by careful
integration of the whole distribution, and the re-evaluation of one
limit at a different c.l. would in principle require a new integration.

{\it ii)} As we have explained, the procedure of fitting in each run
the time dependent spectrum directly from the data, and next using the
inferred spectral function for constructing the Likelihood, implies
that at each new run a slightly different model is being tested. Due
to statistical fluctuations in the spectral fits this becomes a
particularly delicate point when the statistics is low, that is when
detectors with small fiducial volume or when large SN distances are
considered. In these cases one cannot assume a naive scaling of the
results according to the available statistics since, as we will see,
the inferred limits worsen quickly when the number of neutrino events
becomes too small. In all these cases specific runs are required to
infer correctly the sensitivity of the method.

To keep trace of possible non Gaussian effects, for each one of the
cases considered (different SN models, detectors and SN-earth
distances) we have performed a sufficiently large number of tests.
While we have found that in the cases considered non Gaussian effects
never spoil too badly the Gaussian approximation, we stress that this
is as an outcome of our analysis and not an a priori assumption.

\begin{figure}[ht]
\vskip-2mm
\hskip-6mm
\epsfxsize=130mm
 \epsfbox{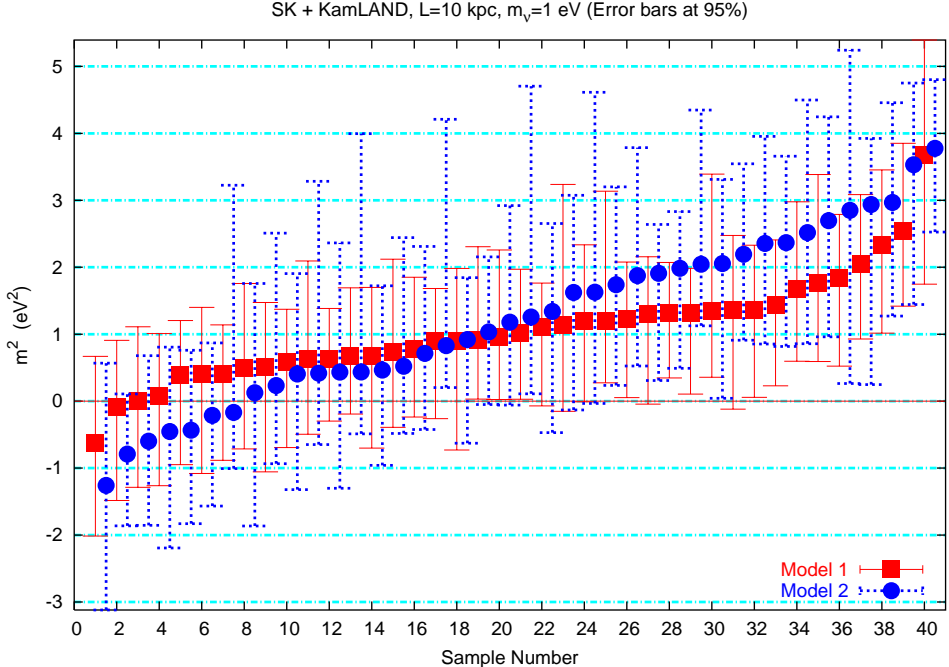}
\vskip-4mm
\caption[]{\baselineskip 12pt Best fit values and 95\% c.l. error bars
for $m_\nu^2$ resulting from 40+40 analysis for the representative case
of a SN at 10~kpc, a neutrino mass of 1 eV, and the combined SK plus
KamLAND data.  The squares and circles refer respectively to SN
simulations performed with model 1 \cite{Woosley:1994ux} and with
model 2 \cite{Raffelt:2003en,Buras:Private2004}. }
\label{fig:bandplot}
\end{figure}

The sensitivity of the method has been tested by analyzing several
neutrino samples, grouped into different ensembles containing about 40
samples each. Each ensemble corresponds to a particular set of input
conditions in the MC code: we vary in turn the SN model (model 1 and
2), the SN-earth distance (5, 10, and 15~kpc) and the detection
parameters (fiducial mass, threshold and energy resolution) specific
for two operative detectors (SK and KamLAND) and two proposed
detectors (HK and LENA) that might be realized in the future.  When
the simulation involves HK, since the very large statistics implies
considerable CPU time, the number of samples in each ensemble is
reduced to 20.  In fig. \ref{fig:bandplot} we present as an example
the best fit values and 95\% c.l. limits on $m^2_\nu$ resulting from
the analysis of 40+40 simulations corresponding to the interesting
case of a SN at 10~kpc, a neutrino mass of 1~eV, and the combined SK
plus KamLAND data. The squares and circles correspond to fits to
neutrino signals generated respectively with SN model 1 and SN model
2.

%

While a set of `band-plots' similar to the ones in
fig.~\ref{fig:bandplot} would be representative of the complete
results of the analysis for each ensemble of MC data, in practice two
types of informations are most relevant: if neutrinos are almost
massless particles, the interesting information is the range of upper
limits that could be set on $m_\nu$, if instead neutrino masses are
sizable, it would be interesting to know which is the smallest mass
value that could be measured with this method.  Accordingly, we have
carried out two kinds of estimates : {\it i)} we have evaluated the
upper limits at 95\% c.l.  that could be put on the neutrino mass from
the analysis of the data, in case $m_\nu$ is too small to produce any
observable delay; {\it ii)} we have estimated for which MC input value
of the mass $m_\nu^{\rm MC}$ a massless neutrino can be rejected with
good confidence (at 95\% c.l.)  in about 50\% of the cases. From the
statistical point of view, the two analysis are carried out as follow:

{\it i) } $m_\nu^{\rm MC}=0$: we evaluate the upper limit $m^2_{\rm up}$ 
by requiring that  
\beq
\int_{-\infty}^{m^2_{\rm up}}
p(m^2_\nu|D,I)\;d\,m^2_\nu  
\simeq 
\int_0^{m^2_{\rm up}}\hat{\cal L}(m^2_\nu|D,I)\;d\,m^2_\nu 
= 95\%, 
\label{eq:muplimit}
\eeq
where, according to (\ref{eq:pdf}), in the second integral the
integration region has been restricted to positive values of
$m^2_\nu$.  Upper limits for $m_\nu$ can be obtained by integrating
the corresponding probability distribution computed from the posterior
probability for $m^2_\nu$: $p(m|D,I)\sim |m|\;p(m^2_\nu|D,I)$.  


{\it ii) } $m_\nu^{\rm MC} > 0$:  for this case we evaluate the 95\% c.l.
lower limits $m^2_{\rm low}$ on the neutrino mass according to 
\beq
\int_{m^2_{\rm low}}^{+\infty}{p(m^2_\nu|D,I)\;d\,m^2_\nu}  
\simeq 
\int_{m^2_{\rm low}}^{+\infty}{
\hat{\cal L}(m^2_\nu|D,I)\;d\,m^2_\nu}  
=  95\%\,,   
\label{eq:mlowlimit}
\eeq
and we search for the MC input mass value $(m_\nu^{\rm MC})^2=
m^2_{\rm min}$ for which the massless hypothesis is rejected in $50\%$
of the cases (i.e. $m^2_{\rm low}>0$ in half of the tests and
$m^2_{\rm low}<0$ in the other half).  This last requirement implies
that in the limit of a very large number of tests $ \langle m^2_{\rm
low} \rangle = \langle m^2_{\nu \rm best\ fit}\rangle - \langle \Delta
m^2_\nu\rangle \to 0$ thus providing an approximate solution for the
condition $ \langle \Delta m^2_\nu\rangle = \langle m^2_{\nu \rm best\
fit}\rangle $ that distinguishes a real measurement from an upper
limit.  In addition, since $\langle m^2_{\nu \rm best\ fit}\rangle \to
m^2_{\rm min} $ this last parameter characterizes the 95\% c.l. width
of the distribution of the best fit masses when the true neutrino mass
has precisely the value $m^2_{\rm min}$, and therefore it contains all
the relevant information.  Note that a result $m^2_{\rm low}>0$ in
(\ref{eq:mlowlimit}) is clearly meaningful only if $\Theta(m^2_\nu)$
that enters the definition of $p(m^2_\nu|D,I)$ is dropped, and the
integration is carried out over the whole real axis (in Bayesian
language, this simply corresponds to a change in the prior).

In the limit in which non Gaussian effects are negligible, the meaning
of $m^2_{\rm up}$ and $m^2_{\rm min}$ is simply that of an estimate of
the (95\%~c.l.) Gaussian width of the distributions, respectively for
the zero mass and for the non-vanishing mass case.  Our results (see
table~\ref{tab:results}) show that for each specific case the average
values of these two quantities to a good approximation are the same,
meaning that the intrinsic widths do not change appreciably when the
test mass is shifted by an amount of the order of 1~eV. This result is
similar to that obtained (for a different range of neutrino masses and
in a somewhat different statistical context) in
refs.~\cite{Beacom:1998ya,Beacom:1998yb}.

The results for the four detectors that we have simulated are
summarized in table~\ref{tab:results}.  The first three rows { a) --
c)} give the results for SK, that is the detector presently in
operation with the largest fiducial volume, for three different
SN-earth distances (5, 10 and 15~kpc).  Using a simple model for the
Galactic rate of star formation \cite{Hartmann:2002ja} we have
estimated that approximately 95\% of the future Galactic SN are likely
to occur between 3~kpc and 17~kpc.  This result is not in disagreement
with a recent study of the Galactic distribution of pulsars, based on
the Parkes Multibeam survey data \cite{Lorimer:2003qc} from which we
have estimated that 93\% of the Galactic core collapse SN occurred
between 2 kpc and 18 kpc.  Therefore, considering also that the
results do not have a strong dependence on the SN-earth distance (see
below) the range of distance 5 -- 15~kpc is sufficient to characterize
the amount of information obtainable from a SN in our Galaxy.

Comparing rows a) and b) in table II, we see that the sensitivity to
the neutrino mass does not vary in going from 10~kpc to 5~kpc.  As is
explained in \cite{Beacom:1998yb}, the approximate independence of the
limits from the SN-earth distance holds for a certain class of
statistical analysis, but might not hold in general. Within the
present approach it holds as long as the total number of events
remains large, and it can be easily understood in terms of naive
scaling of the sensitivity with the square root of the available
statistics. Since the delay in the arrival times increases linearly
with the time of flight, see eq.~(\ref{eq:delay}), the sensitivity to
the neutrino mass square scales with the distance $L$ flew by the
neutrinos, and since the square root of the number of events detected
decreases (geometrically) as $1/L$, the approximate independence of
the sensitivity from the SN-earth distance follows.  However, when we
compare the 10~kpc with the 15~kpc results in row c) we see that this
does not hold anymore.  This is because the efficiency of the method
relies mainly on the large statistics and starts decreasing if the
total number of events is reduced too much.  We see that for model~2
the reduction in the number of events results in a loss of sensitivity
and yields looser limits, while for model~1, whose harder spectrum
still ensures a sufficiently large number of events, this effect is
less important.  Clearly this can be related only to a breakdown of
the scaling law of the sensitivity with the number of events. With a
decrease in the statistics, the uncertainties in the fits to the
spectrum start becoming important since the estimates of the time
dependent spectral functions become not enough accurate. This implies
that the Likelihood does not describe anymore with sufficient
precision the spectral characteristics of the data, and this
represents an additional source of loss of sensitivity.  If the
statistics falls below say, 1000 events, fluctuations in the fits to
the spectrum become too large, and we cannot expect anymore that the
method will perform well.  Luckily, in the case of a large volume
detectors like SK and for a SN in our Galaxy, we are always within the
range in which the efficiency of the method is optimal, but it should
be stressed that its applicability is in fact restricted to these
cases.  For example, the (unlikely) occurrence of another SN in the
nearby Large Magellanic Cloud would yield only about 400 events in SK,
and even in a megaton detector, no more than a couple of dozens of
events can be expected for a SN e.g. in Andromeda. In these cases the
study of the SN signal would require a different method, better suited
for the analysis of sparse data. It is possible that a full
bi-dimensional (in energy and time) maximum Likelihood analysis, in
spite of the fact that it will need to rely on some model-dependent
assumptions about the time dependence of the neutrino spectrum, could
still yield interesting limits.

The second operative detector that we have considered is KamLAND
\cite{Markoff:2003tg}. As we have explained above, our method is not
well suited to analyze the few hundreds of events expected in this
detector.  Therefore, in order to understand how the sensitivity of
scintillator detectors like KamLAND or LVD \cite{Aglietta:2003gi},
that are characterized by a lower threshold, better energy resolution,
but sensibly lower statistics than SK, stands to the sensitivity of a
large volume water \v{C}herencov detector, we have carried out a joint
analysis of the combined SK and KamLAND data. The corresponding
results are given in row { d)}.  Note that such a combined analysis
can be consistently done since SK and KamLAND are located in the same
site, and therefore possible earth-matter effects will modify in
precisely the same way the two neutrino signals (we have also assumed
the same clock for both the detectors).  Comparing the results of the
combined analysis with row { a)} for SK alone, we see that the
sensitivity is completely determined by SK, meaning that the better
energy resolution and lower threshold of KamLAND cannot compete with
the SK much larger statistics.  The results for two of the most
interesting proposed detectors, the megaton water \v{C}herenkov HK
\cite{Nakamura:2002aa} and the multi-kiloton scintillator detector
LENA (Low Energy Neutrino Astrophysics) \cite{LENA} are given in rows
{ e)} and { f)}.  We can see that a megaton detector will be able to
reach a sensitivity about a factor of two better than SK, while a
scintillator detector with a fiducial volume of the order of SK, would
only slightly improve on SK sensitivity.

\begin{table}[t]
\renewcommand{\arraystretch}{1.5}
\begin{tabular}{lc>{\centering}p{2cm}ccc>{\centering}p{2cm}c}
 & \multicolumn{3}{c}{\bf  MODEL 1} &  \phantom{AAAA} 
 & \multicolumn{3}{c}{\bf  MODEL 2} \\ \hline\hline
Detector
& N. events
& $\overline{m}_{\rm up} $ 
& $\sqrt{m_{\rm min}^2}$  
& &  N. events
& $\overline{m}_{\rm up} $ 
& $\sqrt{m_{\rm min}^2}$ \\ 
 & ($\times10^3$) & eV & eV && ($\times10^3$) & eV & eV \\ \hline\hline
%
 a) SK (10 kpc) & 10.0 & $1.1$ & $1.1$ && 5.9 & $1.2$ & $1.2$ \\ 
 b) SK (5 kpc) & 40.0 & $1.2$ & $1.2$ && 23.7 & $1.2$ & $1.2$ \\ 
 c) SK (15 kpc) & 4.4 & $1.4$ & $1.5$ && 2.6 & $1.7$ & $1.8$ \\ 
 d) SK+KL (10 kpc)\qquad\quad & 10.4 & $1.1$ & $1.0$ && 6.1 & $1.2$ & $1.2$ \\ 
%
%
%
\hline
%
%
%
 e) HK (10 kpc) & 170 & $0.5$ & $0.6$ && 100 & $0.6$ & $0.6$ \\ 
 f) LENA (10 kpc) & 12.6 & $1.0$ & $1.0$ && 7.5 & $1.0$ & $1.1$ \\
%
%
%
\hline
%
%
g) SK reference & 9.6 & $0.9$ & $1.0$ && -- & -- & -- \\\hline\hline
%
%
\end{tabular}
\caption{\baselineskip 12pt 
Results for the fits to the neutrino mass at Super-Kamiokande,
Super-Kamiokande plus KamLAND, and at the proposed detectors
Hyper-Kamiokande and LENA. The results for SN model 1 are given in
columns 2-4 and the results for SN model 2 in columns 5-7. The number
of detected neutrino events for different detectors and different
SN-earth distances are given in columns 2 and 5.  The 95\% c.l. upper
limits that could be put on $m_\nu$ for a vanishing MC neutrino mass
are given in columns 3 and 6.
The smaller MC neutrino mass values for which in 50\% of the runs the
95\% c.l.  lower limit $m_{\rm low }$ remains above zero are given in
columns 4 and 7.
}
\label{tab:results}
\end{table}

As we have discussed at the end of the previous section, our
statistical procedure is affected by a certain number of systematic
errors, and these could result in biased estimates of the neutrino
mass values or of the upper limits.  We will now show that the
systematic uncertainty, whether it originates from the effects we have
discussed above or from some other even more subtle mechanism, is
indeed negligible when compared to the statistical fluctuation.  In
table \ref{tab:systematics} we give the average of the best fit values
of the neutrino mass (referred to an input MC mass square of
$1\,$eV$^2$) together with the average of the one standard deviation
statistical errors, for the interesting case of a SN at $10\,$kpc and
the combined SK plus KamLAND data.  The two rows refer to the two SN
models we have been studying in the paper.  Two different sets of 40
signals have been fitted in turn with the two analytical flux models
of eq.~(\ref{eq:fluxmodel1}) and eq.~(\ref{eq:fluxmodel2}) (see also
fig.~\ref{fig:flux-models}).  This test represents an attempt to
estimate possible systematic effects in the procedure, independently
of the particular MC simulation of a SN and of the flux model used for
the fit. We see that each single entry in the second and fourth
columns is completely compatible with the statistical fluctuations
given in the third and fifth columns.  A slight positive bias might be
present in the fits to model 2; however, the statistical error is by
far the dominant source of uncertainty.  Therefore, for all practical
purposes the systematic uncertainties can be neglected, and the
statistical procedure can be considered to a good approximation as
unbiased.
\begin{table}[t] 
\begin{center}
\renewcommand{\arraystretch}{1.5}
\begin{tabular}{p{2cm}>{\centering}p{4cm}>{\centering}c>{\centering}p{4cm}c}
 & \multicolumn{2}{c}{\bf  \qquad Fit with eq. (6)} 
 & \multicolumn{2}{c}{\bf  \qquad Fit with eq. (7)} \\ \hline\hline
Model
&\quad $\langle m^2_{\rm fit}\rangle - (m^{\rm MC}_\nu)^2\quad $
&\quad $\langle\sigma_{\rm stat}\rangle\quad $ 
&\quad $\langle m^2_{\rm fit}\rangle -  (m^{\rm MC}_\nu)^2$\quad
&\quad $\langle\sigma_{\rm stat}\rangle\quad$ \\\hline
SN model 1 
& $+0.05$ & $0.72$
& $-0.19$ & $0.74$ \\ \hline
SN model 2
& $+0.23$ & $0.98$
& $+0.28$ & $0.76$ \\ \hline\hline
\end{tabular}
\caption{\baselineskip 12pt 
The averages of the best-fit values $\langle m^2_{\rm fit}\rangle$ and
of the one standard deviation dispersions of the posterior
probabilities $\langle\sigma_{\rm stat}\rangle$, over two sets of 40
samples generated with SN models 1 and 2, and fitted in turn with the
flux models of eq~(\ref{eq:fluxmodel1}) and eq.~(\ref{eq:fluxmodel2}).
The input MC neutrino mass is $ m_\nu^{\rm MC} =1$ eV, the SN distance
is $10\,$kpc and the samples correspond to the combined data from the
SK and KamLAND detectors.
}
\label{tab:systematics}
\end{center}
\end{table}
%
We believe that the method that we have proposed represents an
improvement with respect to previous techniques, both in sensitivity
and for what concerns the independence from particular astrophysical
assumptions.  It is natural to ask if anything better could be done to
measure neutrino masses from a SN neutrino signal.  In the attempt to
answer this question, we have performed the following test: we have
produced neutrino samples using as inputs to our MC instead than
numerical fluxes and spectra, the simple flux model
(\ref{eq:fluxmodel1}) with a suitable choice of the relevant
parameters. For the time varying spectrum we have used an
$\alpha$-distribution with the (harder) average energy profiles given
in fig.~\ref{fig:models-supernova}b.  We have then performed our usual
set of fits to the neutrino mass (assuming the SK detector) but
we have held the values of the flux shape parameters fixed at the
values used in the MC (only the flux onset parameter $\delta t$ must
be left free to ensure a correct fitting procedure) and we have also
used the same time profile for the spectrum. This simulates the ideal (and
unrealistic) situation where the full time-energy dependence of the
signal at the source is known, and the only relevant free parameter is the
neutrino mass.  The results of this test are given in the last row in
table~\ref{tab:results}, that should be compared with the first
row. We see that only a minor improvement is achieved with respect to
the realistic situation. This allows us to conclude that the
sensitivity to neutrino masses of the detectors presently in operation
is very likely bounded to values not much below 1$\,$eV, and also that
not much sensitivity is lost in the procedure of marginalizing the
nuisance flux parameters.  Future large volume detectors will indeed
reach a sensitivity sizeably better.  However, they will not be
competitive with the next generation of tritium $\beta$-decay
\cite{FutureTritium} and neutrinoless double $\beta$ decay experiments
\cite{Neutrinoless}, or with future high precision cosmological
measurements \cite{Hannestad:2004nb}.


\medskip 
\centerline{\bf Acknowledgments.}
\medskip 

\noindent
We thanks R. Buras for providing us with the numerical results of the Garching
group SN simulation. We acknowledge conversations with Z. Berezhiani and
F. Vissani.  J.I.Z.  acknowledges hospitality from the Laboratori Nazionali di
Frascati (Italy) where part of this research was carried out, and Colciencias
for a scholarship for Doctoral studies.  This work was supported in part by
Colciencias in Colombia under contract 1115-05-13809, and by the Italian
Istituto Nazionale di Fisica Nucleare (INFN).



\bigskip 

\begin{thebibliography}{99}
 \sloppy
%
\bibitem{Atmospheric}
Y. Fukuda {\it et al.}  [Super-Kamiokande Collaboration], 
Phys.\ Rev.\ Lett.\ {\bf 81}, 1562 (1998), hep-ex/9807003;
Y. Fukuda {\it et al.}  [Super-Kamiokande Collaboration], 
Phys.\ Rev.\ Lett.\ {\bf 82}, 2644 (1999), hep-ex/981201;
M. Ambrosio {\it et al.}  [MACRO Collaboration], 
Phys.\ Lett.\ B {\bf 434}, 451 (1998), hep-ex/9807005; 
M. Ambrosio {\it et al.}  [MACRO Collaboration], 
 Phys.\ Lett.\ B {\bf 478}, 5 (2000), hep-ex/0001044; 
M. Sanchez {\it et al.}  [Soudan 2 Collaboration], 
 Phys.\ Rev.\ D {\bf 68}, 113004 (2003), hep-ex/0307069.

\bibitem{Solar}
Y. Fukuda {\it et al.}  [Kamiokande Collaboration], 
Phys.\ Rev.\ Lett.\  {\bf 77}, 1683 (1996);
B. T. Cleveland {\it et al.}, 
Astrophys.\ J.\  {\bf 496}, 505 (1998); 
W. Hampel {\it et al.}  [GALLEX Collaboration], 
Phys.\ Lett.\ B {\bf 447}, 127 (1999);
M. Altmann {\it et al.}  [GNO Collaboration], 
 Phys.\ Lett.\ B {\bf 490}, 16 (2000), hep-ex/0006034; 
S. Fukuda {\it et al.}  [Super-Kamiokande Collaboration], 
 Phys.\ Rev.\ Lett.\  {\bf 86}, 5651 (2001), hep-ex/0103032;
J. N. Abdurashitov {\it et al.}  [SAGE Collaboration], 
 J.\ Exp.\ Theor.\ Phys.\  {\bf 95}, 181 (2002) [Zh.\ Eksp.\ Teor.\ Fiz.\
   {\bf 122}, 211 (2002)],  astro-ph/0204245; 
M. B. Smy {\it et al.}  [Super-Kamiokande Collaboration], 
 Phys.\ Rev.\ D {\bf 69}, 011104 (2004), hep-ex/0309011; 
S. N. Ahmed {\it et al.}  [SNO Collaboration], 
 Phys.\ Rev.\ Lett.\  {\bf 92}, 181301 (2004), nucl-ex/0309004.

\bibitem{Ahmad:2001an}
Q. R. Ahmad {\it et al.}  [SNO Collaboration], 
Phys.\ Rev.\ Lett.\  {\bf 87}, 071301 (2001), nucl-ex/0106015.

\bibitem{KamLANDResults}
K. Eguchi {\it et al.}  [KamLAND Collaboration], 
 Phys.\ Rev.\ Lett.\  {\bf 90}, 021802 (2003), hep-ex/0212021; 
T. Araki {\it et al.}  [KamLAND Collaboration], 
hep-ex/0406035.


\bibitem{Paes:2001nd}
H. Paes and T. J. Weiler, 
 Phys.\ Rev.\ D {\bf 63}, 113015 (2001), hep-ph/0101091.

\bibitem{Bilenky:2002aw}
S. M. Bilenky, C. Giunti, J. A. Grifols and E. Masso, 
 Phys.\ Rept.\  {\bf 379}, 69 (2003), hep-ph/0211462.

\bibitem{Tritium}
J. Bonn {\it et al.}, 
Prog.\ Part.\ Nucl.\ Phys.\  {\bf 48}, 133 (2002); 
V. M. Lobashev {\it et al.}, 
Nucl.\ Phys.\ Proc.\ Suppl.\  {\bf 91}, 280 (2001).

\bibitem{Neutrinoless}
H. V. Klapdor-Kleingrothaus {\it et al.}, 
 Eur.\ Phys.\ J.\ A {\bf 12}, 147 (2001), hep-ph/0103062; 
C. E. Aalseth {\it et al.}  [16EX Collaboration], 
 Phys.\ Rev.\ D {\bf 65}, 092007 (2002), hep-ex/0202026; 
H. V. Klapdor-Kleingrothaus, I. V. Krivosheina, A. Dietz and O. Chkvorets, 
 Phys.\ Lett.\ B {\bf 586}, 198 (2004), hep-ph/0404088.


\bibitem{Hannestad:2004nb}
S. Hannestad,
New\ Jour.\ Phys.\ {\bf 6}, 108 (2004), hep-ph/0404239.

\bibitem{Cosmology}
S. Hannestad, 
 JCAP {\bf 0305}, 004 (2003), astro-ph/0303076; 
O. Elgaroy and O. Lahav, 
JCAP {\bf 0304}, 004 (2003), astro-ph/0303089;
P. Crotty, J. Lesgourgues and S. Pastor, 
 Phys.\ Rev.\ D {\bf 69}, 123007 (2004), hep-ph/0402049.

\bibitem{Blanchard:2003du}
  A.~Blanchard, M.~Douspis, M.~Rowan-Robinson and S.~Sarkar,
  Astron.\ Astrophys.\  {\bf 412}, 35  (2003), astro-ph/0304237.
\bibitem{Beacom:2004yd}
J. F. Beacom, N. F. Bell and S. Dodelson, 
 Phys.\ Rev.\ Lett.\  {\bf 93}, 121302 (2004), astro-ph/0404585.
\bibitem{SupernovaSeminal}
G. T. Zatsepin, 
JETP Lett. 8 (1968) 205, 
[Zh. Eksp. Teor. Fiz. 8, 333 (1968)]; 
S. Pakvasa and K. Tennakone, 
Phys.\ Rev.\ Lett.\  {\bf 28}, 1415 (1972), 
T. Piran, 
Phys.\ Lett.\ B {\bf 102}, 299 (1981), 
Z.~F.~Seidov, 
Astrophysics and Space Science, vol. 81, no. 1-2, Jan. 1982, p. 483-488.

\bibitem{SN1987A:Detection}
K. S. Hirata {\it et al.}, 
Phys.\ Rev.\ D {\bf 38}, 448 (1988); 
J. C. Van Der Velde {\it et al.}  [IMB Collaboration], 
Nucl.\ Instrum.\ Meth.\ A {\bf 264}, 28 (1988); 
E. N. Alekseev, L. N. Alekseeva, I. V. Krivosheina and V. I. Volchenko, 
Phys.\ Lett.\ B {\bf 205}, 209 (1988).

\bibitem{Schramm:1987ra}
D. N. Schramm, 
Comments Nucl.\ Part.\ Phys.\  {\bf 17}, 239 (1987).

\bibitem{SN1987A:Masses}
W. D. Arnett and J. L. Rosner, 
Phys.\ Rev.\ Lett.\  {\bf 58}, 1906 (1987); 
J. N. Bahcall and S. L. Glashow, 
Nature {\bf 326}, 476 (1987);
L. F. Abbott, A. De Rujula and T. P. Walker, 
Nucl.\ Phys.\ B {\bf 299}, 734 (1988).


\bibitem{Loredo:2001rx}
T. J. Loredo and D. Q. Lamb, 
 Phys.\ Rev.\ D {\bf 65}, 063002 (2002), astro-ph/0107260.

\bibitem{Fargion:1981gg}
D. Fargion, 
 Lett.\ Nuovo Cim.\  {\bf 31}, 499 (1981), hep-ph/0110061.

\bibitem{Arnaud:2001gt}
N. Arnaud {\it et al.}, 
 Phys.\ Rev.\ D {\bf 65}, 033010 (2002), hep-ph/0109027.

\bibitem{BH}
J. F. Beacom, R. N. Boyd and A. Mezzacappa, 
 Phys.\ Rev.\ D {\bf 63}, 073011 (2001), astro-ph/0010398;
 Phys.\ Rev.\ Lett.\  {\bf 85}, 3568 (2000), hep-ph/0006015.

\bibitem{Totani:1998nf}
T. Totani, 
 Phys.\ Rev.\ Lett.\  {\bf 80}, 2039 (1998), astro-ph/9801104.




\bibitem{Nardi:2003pr}
E. Nardi and J. I. Zuluaga, 
 Phys.\ Rev.\ D {\bf 69}, 103002 (2004), astro-ph/0306384.

\bibitem{Livermore}
R. Mayle, J. R. Wilson and D. N. Schramm,
Astrophys.\ J.\  {\bf 318}, 288 (1987); 
J. R. Wilson, R. Mayle, S. E. Woosley and T. Weaver,
Annals N.\ Y.\ Acad.\ Sci.\  {\bf 470}, 267 (1986); 
J. R. Wilson and R. W. Mayle,
Phys.\ Rept.\  {\bf 227}, 97 (1993); 
R. W. Mayle, J. R. Wilson and M. Tavani,
Astrophys.\ J.\  {\bf 418} (1993) 398; 
D. S. Miller, J. R. Wilson, R. W. Mayle,
Astrophys.\ J.\  {\bf 415}, 278 (1993).


\bibitem{Burrows:1991kf}
A. Burrows, D. Klein and R. Gandhi, 
Phys.\ Rev.\ D {\bf 45}, 3361 (1992).



\bibitem{Totani:1997vj}
T. Totani, K. Sato, H. E. Dalhed and J. R. Wilson, 
 Astrophys.\ J.\  {\bf 496}, 216 (1998), astro-ph/9710203.


\bibitem{Woosley:1994ux}
S. E. Woosley, J. R. Wilson, G. J. Mathews, R. D. Hoffman and B. S. Meyer, 
Astrophys.\ J.\  {\bf 433}, 229 (1994).

\bibitem{Raffelt:2003en}
G. G. Raffelt, M. T. Keil, R. Buras, H. T. Janka and M. Rampp, 
proceedings of the 4th Workshop on Neutrino Oscillations and their
 Origin (NOON03), Feb. 10-14, 2003, Kanazawa, Japan, edited by
 Y. Suzuki, M. Nakahata, Y. Itow, M. Shiozawa \& Y. Obayashi (World
 Scientific, Singapore, 2004), pp. 380-387, astro-ph/0303226.

\bibitem{Janka:1989aa}
H. T. Janka and W. Hillebrandt, 
Astron.\ \&.\ Atrophys.\  {\bf 224}, 49 (1989).

 \bibitem{Nardi:2004ms}
E.~Nardi,
to appear in the proceedings of the 10th Marcel Grossmann Meeting on
Recent Developments in Theoretical and Experimental General
Relativity, Gravitation and Relativistic Field Theories (MG X MMIII),
Rio de Janeiro, Brazil, 20-26 Jul 2003, astro-ph/0401624.


\bibitem{Buras:Private2004}
R. Buras, Private Communication.


\bibitem{Buras:2002wt}
R.~Buras, H.~T.~Janka, M.~T.~Keil, G.~G.~Raffelt and M.~Rampp,
Astrophys.\ J.\  {\bf 587}, 320 (2003), astro-ph/0205006.

\bibitem{Keil:2002in}
M. T. Keil, G. G. Raffelt and H. T. Janka,
 Astrophys.\ J.\  {\bf 590}, 971 (2003), astro-ph/0208035; 
M. T. Keil, Ph. D. Dissertation (2003):  
``Supernova neutrino spectra and applications to flavor oscillations'', 
 astro-ph/0308228.

\bibitem{Dighe:2003jg}
A. S. Dighe, M. T. Keil and G. G. Raffelt, 
 JCAP {\bf 0306}, 006 (2003), hep-ph/0304150.

\bibitem{Nakamura:2002aa}
K. Nakamura, 
talk given at the conference ``Neutrino and Implications for Physics Beyond the
Standard Model'', 
Stony Brook, NY (2002) [http://insti.physics.sunysb.edu/itp/conf/
neutrino/talks/nakamura.pdf].

\bibitem{Markoff:2003tg}
D.~M.~Markoff  [KamLAND Collaboration],
J.\ Phys.\ G {\bf 29}, 1481 (2003).


\bibitem{LENA}
F.~von~Feilitzsch, L.~Oberauer and W.~Potzel,
prepared for the Eighth International Workshop On Topics In Astroparticle
And Underground Physics, TAUP 2003, September 5 - 9, 2003, University
of Washington, Seattle, Washington, 
[http://www.int.washington.edu/talks/
WorkShops/ TAUP03/Parallel/People/Oberauer\_L/LENA-Oberauer.pdf];
L. Oberauer, 
 Mod.\ Phys.\ Lett.\ A {\bf 19}, 337 (2004), hep-ph/0402162.


 \bibitem{Nardi:2004uz}
E.~Nardi,
to appear in the proceedings of the 5th Latin American Simposium 
on High Energy Physics (SILAFAE-V), Lima, Peru,  July 12-17 2004, 
hep-ph/0412024.

\bibitem{FutureTritium}
A. Osipowicz {\it et al.}  [KATRIN Collaboration], 
 hep-ex/0109033; 
C. Weinheimer  [KATRIN Collaboration],
Prog.\ Part.\ Nucl.\ Phys.\  {\bf 48}, 141 (2002).


\bibitem{Cremonesi:2002is}
O. Cremonesi,
 Nucl.\ Phys.\ Proc.\ Suppl.\  {\bf 118}, 287 (2003), hep-ex/0210007.

\bibitem{Strumia:2003zx}
A. Strumia and F. Vissani,
 Phys.\ Lett.\ B {\bf 564}, 42 (2003), astro-ph/0302055.


\bibitem{Vogel:1999zy}
  P.~Vogel and J.~F.~Beacom,
  Phys.\ Rev.\ D {\bf 60} (1999) 053003
  [arXiv:hep-ph/9903554].

\bibitem{DAgostini:2003qr}
G. D' Agostini,
 Rept.\ Prog.\ Phys.\  {\bf 66}, 1383 (2003), physics/0304102.

\bibitem{Buras:2003sn}
R. Buras, M. Rampp, H. T. Janka and K. Kifonidis,
 Phys.\ Rev.\ Lett.\  {\bf 90}, 241101 (2003), astro-ph/0303171.

\bibitem{GarchingCode}
M. Rampp,
Ph. D. Dissertation : ``Radiation Hydrodynamics with Neutrinos: Stellar 
Core Collapse and the Explosion Mechanism of Type 
II Supernovae'',  
[http://tumb1.biblio.tu-muenchen.de/publ/diss/ph/2000/rampp.pdf], (2000);
M. Rampp and H. T. Janka,
 Astron.\ Astrophys.\  {\bf 396}, 361 (2002), astro-ph/0203101; 
Astrophys.\ J.\  {\bf 539}, L33 (2000), astro-ph/0005438.

\bibitem{Janka:2004tt}
H. T. Janka, R. Buras, F. S. Kitaura Joyanes, A. Marek and M. Rampp, 
 astro-ph/0405289.



\bibitem{Maltoni:2004ei}
M.~Maltoni, T.~Schwetz, M.~A.~Tortola and J.~W.~F.~Valle,
New\ Jour.\ Phys.\ {\bf 6}, 122 (2004), hep-ph/0405172.

\bibitem{Bahcall:2004ut}
J. N. Bahcall, M. C. Gonzalez-Garcia and C. Pena-Garay,
 JHEP {\bf 0408}, 016 (2004), hep-ph/0406294.


\bibitem{Dighe:1999bi}
A. S. Dighe and A. Y. Smirnov,
 Phys.\ Rev.\ D {\bf 62}, 033007 (2000), hep-ph/9907423.


\bibitem{Lunardini:2001pb}
C. Lunardini and A. Y. Smirnov,
 Nucl.\ Phys.\ B {\bf 616}, 307 (2001), hep-ph/0106149.


\bibitem{Nakahata:1998pz}
M. Nakahata {\it et al.}  [Super-Kamiokande Collaboration],
 Nucl.\ Instrum.\ Meth.\ A {\bf 421}, 113 (1999), hep-ex/9807027.


\bibitem{Beacom:2003nk}
J. F. Beacom and M. R. Vagins,
Phys.\ Rev.\ Lett.\  {\bf 93}, 171101 (2004), hep-ph/0309300.

\bibitem{Virtue:2001mz}
C.~J.~Virtue  [SNO Collaboration],
Nucl.\ Phys.\ Proc.\ Suppl.\  {\bf 100}, 326 (2001), 
astro-ph/0103324.


\bibitem{Aharmim:2004uf}
B.~Aharmim {\it et al.}  [SNO Collaboration],
Phys.\ Rev.\ D {\bf 70}, 093014 (2004), hep-ex/0407029.

\bibitem{Iwamoto:2003aa}
T. Iwamoto,
Ph. D. Dissertation (2003): ``Measuring Reactor Anti-Neutrino Dissapearance 
in KamLAND'',
[http://www.awa.tohoku.ac.jp/KamLAND/articles/PhD/PhD-iwamoto.pdf].  

\bibitem{Jung:1999jq}
C. K. Jung,
to appear in the proceedings of International Workshop on Next
Generation Nucleon Decay and Neutrino Detector (NNN 99), Stony Brook,
New York, 23-25 Sep 1999, hep-ex/0005046.


\bibitem{Beacom:1998ya}
  J.~F.~Beacom and P.~Vogel,
  Phys.\ Rev.\ D {\bf 58} (1998) 053010
  [arXiv:hep-ph/9802424].

\bibitem{Beacom:1998yb}
  J.~F.~Beacom and P.~Vogel,
  Phys.\ Rev.\ D {\bf 58} (1998) 093012
  [arXiv:hep-ph/9806311].


\bibitem{Hartmann:2002ja}
  D.~H.~Hartmann, K.~Kretschmer and R.~Diehl,
  arXiv:astro-ph/0205110.

\bibitem{Lorimer:2003qc} D.~R.~Lorimer, 
ASP Conference Proceedings of IAU Symposium 218, ``Young Neutron Stars and
  their Environments", Sydney, Australia, 14-17 Jul 2003, eds.
  F. Camilo and B. M. Gaensler, Vol. 218, (2004), astro-ph/0308501;
  N.~Vranesevic {\it et al.},
  Astrophys.\ J.\  {\bf 617} (2004) L139, 
  astro-ph/0310201.
\bibitem{Aglietta:2003gi}
M.~Aglietta {\it et al.}  [LVD Collaboration],
``10 years search for neutrino bursts with LVD'',
prepared for 28th International Cosmic Ray Conferences (ICRC 2003),
Tsukuba, Japan, 31 Jul - 7 Aug 2003,  
[http://www.slac.stanford.edu/spires/find/hep/www?irn=5876737]. 






\end{thebibliography}
\end{document}